\documentclass[dvipsnames,twocolumn]{aastex63}

\newcommand{\vecb}[1]{\mathbf{#1}}

\newcommand{\HI}{21\,cm\xspace}
\newcommand{\model}[1]{\textsc{#1}}
\newcommand{\code}[1]{\textsc{#1}}

\usepackage{amsmath}
\usepackage{mathtools}
\usepackage{tabularx}
\usepackage{appendix}
\usepackage[capitalise]{cleveref}
\usepackage{multirow}
\usepackage{etoolbox}
\usepackage{xspace}

\newmuskip\pFqmuskip

\newcommand*\pFq[6][8]{%
  \begingroup 
  \pFqmuskip=#1mu\relax
  \mathchardef\normalcomma=\mathcode`,
  \mathcode`\,=\string"8000
  \begingroup\lccode`\~=`\,
  \lowercase{\endgroup\let~}\pFqcomma
  {}_{#2}F_{#3}{\left(\left.\genfrac..{0pt}{}{#4}{#5}\right|#6\right)}%
  \endgroup
}
\newcommand{\pFqcomma}{{\normalcomma}\mskip\pFqmuskip}

\DeclareMathOperator{\sinc}{sinc}

\received{December 30, 2020}
\revised{September 6, 2021}
\revised{December 21, 2021}
\accepted{\today}
\submitjournal{ApJS}

\shorttitle{RIME Solutions for Diffuse Emission}
\shortauthors{Lanman, Murray \& Jacobs}

\begin{document}


\title{Validation Solutions to the Full-Sky Radio Interferometry Measurement Equation for Diffuse Emission}

\correspondingauthor{Adam E. Lanman}
\email{adam.lanman@mcgill.ca}
\author[0000-0003-2116-3573]{Adam E. Lanman}
\affiliation{Department of Physics, McGill University, 3600 rue University, Montr\'{e}al, QC H3A 2T8, Canada}
\affiliation{Brown University, Department of Physics, Providence, RI, 02912, USA}

\author[0000-0003-3059-3823]{Steven G. Murray}
\affiliation{Arizona State University, School of Earth and Space Exploration, Tempe, AZ, 85008, USA}

\author[0000-0002-0917-2269]{Daniel C. Jacobs}
\affiliation{Arizona State University, School of Earth and Space Exploration, Tempe, AZ, 85008, USA}



\begin{abstract}

Low-frequency radio observatories are reaching unprecedented levels of sensitivity in an effort to detect the \HI signal from the Cosmic Dawn. High precision is needed because the expected signal is overwhelmed by foreground contamination, largely from so-called \emph{diffuse} emission -- a non-localized glow comprising Galactic synchrotron emission and radio galaxies. The impact of this diffuse emission on observations may be better understood through detailed simulations, which evaluate the Radio Interferometry Measurement Equation (RIME) for a given instrument and sky model. Evaluating the RIME involves carrying out an integral over the full sky, which is naturally discretized for point sources but must be approximated for diffuse emission. The choice of integration scheme can introduce errors that must be understood and isolated from the instrumental effects under study. In this paper, we present several analytically-defined patterns of unpolarized diffuse sky emission for which the RIME integral is manageable, yielding closed-form or series visibility functions. We demonstrate the usefulness of these RIME solutions for validation by comparing them to simulated data, and show that the remaining differences behave as expected with varied sky resolution and baseline orientation and length.
\end{abstract}




\section{Introduction} \label{sec:intro}
Detection and characterization of \HI\ emission during and prior to the Epoch of Reionization (EoR) has motivated substantial investment in new low-frequency radio telescopes. The IGM during the EoR is expected to comprise large ($\sim100$~Mpc scale) structures, which will be detectable as a faint, diffuse field of \HI emission and absorption. This has motivated the design of wide-field, compact interferometer arrays that are very sensitive to low-frequency diffuse radio emission, but are also very sensitive to diffuse foregrounds -- Galactic synchrotron emission and the collective contributions of unresolved radio galaxies. These foreground components appear strongest on short baselines and are visible in all directions and on all angular scales \citep{haslam1982,costa2008, Kim2018,thyagarajan2016,Presley:2015}. Although several methods of avoiding or subtracting this diffuse foreground power have been developed, \HI experimental results are still limited by residual foreground power. 

Simulations of expected instrumental output are used for instrument  design \citep{ewall-wice2016,thyagarajan2016}, pipeline verification \citep{patil2016, aguirre2021_arxiv}, and calibration~\citep{li_first_2019}. Simulating a 100~K-1000~K smooth spectrum foreground against a 10~mK background requires that any visibility errors be smaller than one part in 10,000. The most important metric in all of these cases is the amount of spectrally smooth foreground power coupled into power spectrum modes which should otherwise be dominated by \HI background. When used in calibration, errors in the simulations can appear to produce foreground bias when none is there \citep{barry2016}. Errors in simulated output are not commonly reported on in the literature but are commonly found during their development. Errors found in comparison with data run the risk of experimenter bias. Differences between nominally independent simulators tell us about the simulator precision but doesn't tell us which one is right.  In our development of the simulators used to validate the recent HERA pipeline \citep{aguirre_validation_2021} we found a need for a way to check diffuse and \HI forward model simulators quickly and at better than 1 part in 10,000 precision.



The state of the art in interferometer simulator precision is probably best demonstrated by the accuracy of foreground subtraction methods \citep{li_first_2019,mertens_improved_2020}. In these cases point sources formed the basis of the model. The accuracy of simulated point-source visibilities, discounting code errors, is limited by the quality of the instrument and catalogs. It is relatively straightforward to confirm in an analytic way that a point source simulator code produces no unwanted spectral structure.  

Diffuse galactic emission, however, is a bright foreground at large scales and the \HI background is intrinsically diffuse.  The galaxy has diffuse power on all spatial scales with both a isotropic component and the bright spatially distinct galactic plane spans the entire sky. The spatial \HI power spectrum follows a power law, peaking at an angular scale of several degrees \citep{furlanetto_cosmology_2006}. Simulation of these elements requires precision evaluation of the RIME, the \emph{Radio Interferometric Measurement Equation}, integral \citep{thompson_interferometry_2017}. The visibility $V$ is given by an integral of a primary beam function $A$, a surface brightness function $T$, and a fringe term with period set by baseline vector $\vecb{b}$ and frequency $\nu$n and time $t$:
\begin{equation}
V(\vecb{b}, \nu) = \int\limits_{\text{sky}} A(\hat{s}, \nu) T(\hat{s}, \nu) e^{-2 \pi i \vecb{b} \cdot (\hat{s} - \hat{s}_0) \nu / c} d\Omega
\label{eqn:rime_scalar_full}
\end{equation}
In \cref{eqn:rime_scalar_full}, $\hat{s}$ is a unit vector pointing toward a position on the sky, $\hat{s}_0$ points to a reference phase center, and the integral is carried out over the unit hemisphere.\footnote{This form of the RIME explicitly ignores polarization, which is an important factor for high-precision interferometry. In a full treatment with polarization, \cref{eqn:rime_scalar_full} will be made up of vectors \citep[see e.g.][]{smirnov_revisiting_2011}. The integral is carried out on each component separately, and so working with the scalar form of the RIME does not limit the applicability of our results.} In typical visibility simulations, $T$ (and sometimes $A$) will be derived from some dataset with a finite spatial resolution, while the fringe term is continuous across the sky. Numerically evaluating such an integral can be challenging, especially one with an oscillating term like the exponential fringe.

There are several popular approaches to carrying out this integral, such as by treating map pixels as point sources (e.g., \code{OSKAR}~\cite{kloeckner_ska_2010,mort_oskar_2010}, \code{wsclean}~\cite{arras_efficient_2021}, and \code{Meqtrees}\footnote{\code{Meqtrees} is capable of simulating resolved sources as Gaussians, shapelets, or as point-source pixels, but we can find no reference to using it for widefield diffuse simulation.} \cite{noordam_meqtrees_2010}), or as small Gaussians (e.g., \code{PRISim} \cite{thyagarajan_nithyanandanprisim_2020}), or transforming the RIME to spherical harmonic space (e.g., \code{RIMEz} \cite{aguirre_validation_2021}). Carefully-weighted quadrature methods may be used at the expense of having to interpolate sky and beam models to the quadrature nodes. Importantly, in all cases, some discretization is required in order to evaluate the RIME.

A visibility simulator must be validated to ensure that its accuracy is limited only by the quality of the instrument and sky models used, and not on the integration scheme. There are two main ways one might expect a diffuse integration scheme to produce inaccurate results: firstly it might be that the specific discretization chosen is too coarse (e.g. too few spherical harmonics, or pixels that are too large) to be accurate, and secondly it may be that even with arbitrary resolution some fundamental aspect of the scheme produces inaccuracies (code bugs and poor algorithm choices fall into this category).\footnote{A softer in-between category of inaccuracy would be algorithms that converge to an accurate answer very slowly as their resolution increases, as would be the case for simulating a sparse point-source sky via discretization into healpix pixels.}

How does one appropriately validate that a given simulator is free of both of these kinds of errors?
One approach would be to simulate a sky model\footnote{Throughout this paper, phrases like ``simulate a sky model'' may be interpreted to mean ``generate simulated visibilities for a given sky and instrument model.'' This should not be confused with the task of modeling sky emission.} at progressively higher spatial resolutions (in whatever basis one chooses to discretize in) and check that the resulting visibilities converge on a particular value. 
This method lets us infer resolution required to achieve a particular precision. 
However, the ``many sources'' discretization test has a number of drawbacks.
Most importantly, one has no way of telling whether the simulator is converging to the correct answer -- it may be precise, but not accurate. Comparing to a second simulator would also face the same problem. We could compare simulation products to data, but then we cannot untangle errors from our integration scheme from errors in sky and instrument modeling. Also, any method where precision is achieved by adding more items to the sum is subject to accumulation of numerical machine precision errors which grow as points are added. This could potentially effect the accuracy of a convergence. Such a method is also inefficient, potentially requiring significant computing resources to converge.
Lastly, and perhaps most pragmatically, it doesn't offer any insight into the specific causes of inaccuracy.

An ideal validation method would be an independent technique to produce \textit{arbitrary precision} visibilities of sky power distributions relevant to a diffuse integrator.
%
%
%
%
%
%
A sound method along these lines
is to find an integrand $I(\hat{s}) = A(\hat{s})
T(\hat{s})$ which makes the integral soluble,
just like one can do for a point source. We have found solutions for a selection of integrands which are similar in their properties to the diffuse, all-sky, nature of the galaxy and 21cm background.  Throughout this paper, we will refer to these functions $I(\hat{s})$ as \emph{patterns} and the corresponding visibility function $V(\vecb{b})$ as \emph{solutions}. By comparing simulated visibilities to an evaluated solution, one can directly calculate accuracy of a simulator and explore how it depends on the discretization scheme.

Such ``solutions'' don't need to be closed-form, but should be easily computable to arbitrary precision (e.g. a quickly converging infinite sum does not negate any of the benefits of the method). Series solutions still have the benefit that their precision is independent of the properties of any simulator they're used to validate, and they can be known to converge to the correct values. Comparing simulations to solutions for carefully-chosen test patterns thus overcomes the downsides associated with validation via convergence testing.

Under the assumption of narrow field of view, the integral of \cref{eqn:rime_scalar_full} reduces to a 2D Fourier transform, for which dozens of exact pattern/solution pairs are known. These may be sufficient for validating simulations of resolved extended sources, but are inadequate for testing simulators that aim to accurately simulate horizon-to-horizon diffuse structure. The full RIME includes the complications associated with a sharp cutoff at the horizon and projection of the fringe at low altitudes. Several effects associated with the horizon have been found to be important to 21 cm science, many discovered from instrument simulations, and so verifying that simulators can handle the horizon correctly will be important to the future use of instrument simulation in this field.

In this paper, we describe several analytically-defined patterns of full-sky diffuse emission for which a closed-form or converging series solution can be found. These solutions allow RIME integration schemes to be tested for errors vs baseline lengths and orientations for situations with objects extending across the visible sky to the horizon, non-coplanar baselines, large sources far from zenith, large features with sharp edges, and rotational asymmetry.\footnote{Note that no single pattern  provides \emph{all} of these features at once, but all of these features are represented in the collection.}  We demonstrate how these patterns and their solutions may be used for validation by comparing them to a basic, brute-force diffuse simulation code called \code{healvis} that approximates pixels (sampled on a \code{healpix} grid) as point sources. A Python implementation of these patterns and their solutions is publicly available\footnote{\url{github.com/aelanman/analytic_diffuse}} and a subset of them have been implemented as unit tests in the \code{pyuvsim}\footnote{\url{https://github.com/RadioAstronomySoftwareGroup/pyuvsim}} validation simulator.

The outline of the paper is as follows; in \S\ref{sec:definitions} we define our general conventions for the RIME and sky co-ordinates.
In \S\ref{sec:analytic_solns} we derive a set of exact solutions to the RIME for analytic diffuse sky models, using two different approaches.
In \S\ref{sec:simulator} we describe a basic discretization that performs the integral in \cref{eqn:rime_scalar_full} as a Riemann sum, and derive expressions giving approximate errors due to this scheme. 
In \S\ref{sec:comparison}, we compare this simulator to the analytic predictions, noting trends in the magnitude of errors with various input parameters (such as sky model resolution).
We summarize our conclusions in \S\ref{sec:conclusions}.

\section{General Definitions}
\label{sec:definitions}

In line with \cite{thompson_interferometry_2017}, we rewrite the RIME integral, \cref{eqn:rime_scalar_full}, in terms of direction cosines:
\begin{equation}
    V(u,v,w) = \int\limits_{(l,m) \in D_1} \frac{d^2\mathbf{l}}{n}\ A(\mathbf{l}) T(\mathbf{l}) e^{-2\pi i [\mathbf{u}\cdot\mathbf{l} + w(1 - n)]},
\label{eq:vis_lmn}
\end{equation}
where $\vecb{u} \equiv (u,v)$ is the baseline vector projected to the tangent plane of the phase center, in units of wavelength, $\vecb{l} \equiv (l,m)$ is a position on the sky in the $u and v$ directions, $n = \sqrt{1 - \vecb{l}}^2$, and $w$ is the component of the baseline vector toward the phase centre. $D_1$ denotes the unit disc, which comprises all points $(l,m)$ such that $l^2 + m^2 < 1$. The $n$ in the denominator comes from the Jacobian of the transformation from angular to cosine coordinates. \Cref{eq:vis_lmn} is entirely equivalent to \cref{eqn:rime_scalar_full}.

We separate the $w$ term from the baseline vector because most of our solutions assume antennas are coplanar, meaning a phase centre can be chosen so that all baselines have $w=0$. This is typically the case for compact \HI arrays.  For coplanar antennas, off-zenith phase centres can be related to a zenith phase centre by a simple phase term which falls outside of the RIME integral, and so has no bearing on the problem of numerically integrating the RIME. For arrays with non-coplanar baselines, however, the factor of $w (1 - n)$ in the fringe will affect the RIME integral. We present a specific case of non-zero $w$ which can be calculated analytically.

Since we can factor the term $\exp(-2\pi i w)$ out of the integrand, we will omit this from the RIME for the rest of the paper for simplified notation. Doing so is equivalent to assuming a different phase center for each baseline, at a position orthogonal to the baseline vector. Trivially rephasing the visibilities ensures a common phase center for all baselines. Most importantly, since this term factors out of the integrand, removing it does not affect the problem of numerically evaluating the integral.


For completeness, we note the following identities that are true under these definitions, in which $\phi$ is the zenith angle (i.e. angle from zenith towards horizon) and $\theta$ is the angle from the $m=0$ line around the North celestial pole (increasing towards positive $m$):

\begin{equation}
\begin{aligned}
    l &= \cos \theta \sin\phi, \ \ \ &m &= \sin \theta \sin\phi, \\
    r^2 &= l^2 + m^2 = \sin^2 \phi \ \ \
    &\phi &= \arcsin r, \\  
    \theta &= \arctan{m/l}, \ \ \ &n &= \cos\phi \\
    d \Omega &= d^2\vec{l} / n, \ \ \ &\hat{s} &= (l,m,n)
\end{aligned}
    \label{eq:definitions}
\end{equation}
Note that we use $\vecb{l}$ to represent the 2-vector $(l,m)$ and $\hat{s}$ to represent the Cartesian 3D unit-vector $(l,m,n)$.

Throughout this paper, $A$ and $T$ are entirely interchangeable (i.e. we always consider a single ``beam-weighted sky''), and thus for notational clarity we rewrite $AT \implies I$ here on. That is,
\begin{equation}
    V(u,v,w) = \int \frac{d^2\mathbf{l}}{n}\ I(\mathbf{l}) e^{-2\pi i [\mathbf{u}\cdot\mathbf{l} - w n]}.
\label{eq:rime_simple}
\end{equation}

We note that the factor of $1/n$ introduces an apparent coordinate singularity at the horizon, where $|\vecb{l}| \rightarrow 1$ and so $n = \sqrt{1-|\vecb{l}|^2} \rightarrow 0$. The $(l,m,n)$ coordinates are a cartesian system, with the $(l,m)$ plane equivalent to the tangent plane at zenith and $n$ representing height below the tangent plane. If we think of the integral measure $d^2\vecb{l} = dl dm$ as an infinitesimal square in the $(l,m)$ plane, the factor of $1/n$ comes from projecting its area onto the unit sphere.

We stress that although $1/n$ blows up near $|\vecb{l}| = 1$, the integral itself converges for any physically-plausible (i.e. non-singular) sky pattern (whether in (l,m,n) or spherical or any other co-ordinates). A uniform-amplitude sky pattern ($I(l,m)=1$) on a zero-length baseline ($\vecb{u}$) reduces \cref{eq:rime_simple} to:
\begin{align*}
V(0) &= \int\limits_{(l,m) \in D_1} \frac{dl dm}{\sqrt{1 - l^2 - m^2}} \\
&= 2 \pi \int\limits_0^1 \frac{dx}{\sqrt{1-x^2}}\\
&= 2 \pi (\pi/2).
\end{align*}
Hence, as long as the sky pattern is non-singular, the integral is convergent.

\section{Analytic Tests} \label{sec:analytic_solns}

In this section, we derive a suite of solutions to the visibility equation, \cref{eq:rime_simple}. We want patterns that can be used to test a simulator's ability to handle different features of diffuse skies, including:
\begin{enumerate}
\item \textbf{Significant brightness near the horizon.} 
    Many instruments have wide primary beams with sensitivity near the horizon. Brightness near the horizon is especially prone to contaminating \HI power spectrum measurements \citep{thyagarajan_foregrounds_2015, lanman_quantifying_2020}.
\item \textbf{Rapid spatial fluctuations.}
	Diffuse galactic emission is concentrated along the Galactic plane, and exhibits structure on all scales probed by most surveys, between 1$^\circ$ -- 180$^\circ$.
\item \textbf{Brightness not centered on the zenith.}
	It is convenient to solve for solutions with brightness centered on the zenith, but finding solutions that break this assumption will provide patterns closer to reality.
	Patterns symmetric about the zenith will yield visibilities with zero imaginary component, which can hide conjugation errors.
\item \textbf{Rotational asymmetry.}
	Breaking rotational symmetry can reveal issues associated with baseline orientation.
\item \textbf{Non-coplanar antennas ($w \neq 0$).}
	Non-zero $w$ can introduce additional phase which is large for wide-field observations, which should be tested. Further, letting $w \neq 0$ breaks the circular symmetry of the horizon for the fringe term.
\end{enumerate}
Due to the difficulty of carrying out the integral in \cref{eq:rime_simple}, only one of our solutions has the last feature ($w \neq 0$). In integrating widefield diffuse skies, the most significant complications come from the sharp cutoff at the horizon and from the factor of $1/n$ in the integrand. These features are still present even when $w = 0$.

\Cref{tab:summary_of_models} summarizes the patterns derived in this section and their solutions, as well as their properties and the motivation for including each. There are three main mathematical routes we take to these solutions. In the first, we consider patterns which are azimuthally symmetric around the zenith and include a factor of $n$ to cancel out the $1/n$. In this case, the integral can be reduced to 1 dimension as a Hankel transform, which neatly yields a few simple solutions. These have the unfortunate quality, however, of having very little response near the horizon. In the second approach, we take the horizon cutoff and the $1/n$ factor as a separate function inside the integrand (with zero response outside) which allows us to take the integral bounds to infinity and treat the RIME exactly as a 2D Fourier transform. In this case, the horizon itself can be Fourier transformed and convolved with the FT of the pattern. This second approach provides some insight into the overall effect of a sharp horizon cutoff which may be of value for designing new simulators. For the last approach we define functions to be zero outside of a square region inscribed in the horizon. This allows the integrand to be separated into two orthogonal axes, allowing for patterns with more complicated angular dependence to be constructed.

Several special functions and integral identities are used throughout this section. We define these in App. \ref{app:special_funcs}.

\begin{table*}
    \begin{tabularx}{\linewidth}{ llll>{\raggedright\arraybackslash}p{5.0cm} }
    \hline
    \textbf{Model} & \textbf{Sky Model}, $I(\phi)$ &  \textbf{Solution} & \textbf{Ref.} & \textbf{Rationale} \\
    \hline
    \model{cosza} &  $\cos\phi$ & Hankel & \S\ref{sec:hankel:gencosine}, \cref{eq:cossky}  & Simple. \\
    \model{gencos} &  $\cos^{n}\phi$ & Hankel & \S\ref{sec:hankel:gencosine}, \cref{eq:gencosine} & Arbitrary width, zero at horizon.  \\
    \model{polydome} &  $[1 - \sin^{2n}(\phi)] \cos\phi$ & Hankel & \S\ref{sec:hankel:polydome}, \cref{eq:polydome} & Arbitrary width, zero at horizon.  \\
    \model{quaddome} &  $[1 - \sin^{2}(\phi)] \cos\phi$ &  Hankel & \S\ref{sec:hankel:polydome}, \cref{eq:quaddome} & Simplest \model{polydome} \\
      \model{projgauss}   &  $\exp(-\sin^2\phi / \sigma^2) \cos\phi$ & Hankel & \S\ref{sec:hankel:projgauss}, \cref{eq:projgauss:largesig,eq:projgauss:smallsig} & Widely used, arbitrary width\\
    \model{monopole} &  $I_0$ & Conv. & \S\ref{sec:convolutions:monopole}, \cref{eq:monopole}  & Strong at horizon, simplest\\
    \model{gauss} &  $\exp(-\sin^2\phi/2\sigma^2)$ & Conv. & \S\ref{sec:convolutions:gaussian}, \cref{eq:gaussian_sine_expansion}, \cref{eq:gaussian_expansion} & Widely used, arbitrary width \\
    \model{shiftgauss}$^\dagger$ &  $\exp(-(\sin\phi - \sin\phi_0)^2 /2\sigma^2)$ & Conv. & \S\ref{sec:convolutions:gaussian}, \cref{eqn:dispgaussian_largea}, \cref{eqn:dispgaussian_smalla} & Test baseline orientation, imaginary components\\
    \model{xysincs}$^*$ &  \cref{eq:separable_I} & Separable & \S\ref{sec:separable}, \cref{eq:xysinc}  & Test baseline orientation, different size scales\\
    \hline
\end{tabularx}
\caption{Summary of test patterns that have closed-form or convergent-expansion solutions and their labels for use throughout the rest of the paper. Hankel solutions are direct integrations of \cref{eq:hankel_solution}. Convolutional solutions are obtained using \cref{eq:conv:axisymmetric} for symmetric patterns and \cref{eqn:convolutional:offzen} for off-zenith patterns. All patterns are circularly symmetric about the zenith except when marked with $\dagger$ (symmetric but shifted from zenith) or $*$ (separable in $l,m$). }
\label{tab:summary_of_models}
\end{table*}





\subsection{Hankel Transform}
\label{sec:hankel}

For this approach, we take $w=0$ and $I(l,m) = I(r)$ where $r = |\vecb{l}|=\sin\phi$ (as in \cref{eq:definitions}). These patterns are thus azimuthally symmetric, and we're only integrating them for coplanar baselines.

With these assumptions, the RIME is equivalent to a Hankel transform (cf. \cref{eq:fourier_to_hankel}):
\begin{equation}
    V(u) = 2\pi \int_0^1 \frac{r I(r)}{\sqrt{1 - r^2}} J_0(2\pi u r)\ dr,
\label{eqn:hankel0}
\end{equation}
where $J_0$ is the zeroth-order Bessel function of the first kind (see App.~\ref{app:special:bessel}) and the integration limits reflect the horizon cutoff. The resulting visibility is also azimuthally symmetric, depending only on the baseline length in wavelengths $u = |\vecb{u}|$.

To remove the factor of $1/\sqrt{1 - r^2} = 1/n$ in \cref{eqn:hankel0}, we absorb it into the brightness function $I$. The consequence of this is that all solutions with this assumption have a factor of $\cos \phi$ in them. We thus proceed by defining the ``projected'' sky intensity $I'$ as the true intensity divided by the cosine of the zenith angle:
\begin{equation}
    I'(r) = \frac{I(r)}{\cos\phi} = \frac{I(r)}{\sqrt{1 - r^2}}.
\end{equation}
The RIME then takes the form,
\begin{equation}
    V(u) = 2\pi \int_0^1 r I'(r) J_0(2\pi u r)\ dr.
    \label{eq:hankel_solution}
\end{equation}

In summary --  this first approach will yield only azimuthally-symmetric patterns which have a $\cos \phi$ dependence, and so won't have any strong response near the horizon or dependence on baseline orientation. However, this can yield some patterns that are simple to implement and that feature some useful parametrization. In later sections we will consider more complicated patterns, but these assumptions make a good starting point for some simple diffuse RIME solutions.

%
%

\subsubsection{Cosine Sky}
\label{sec:hankel:gencosine}
Consider a projected brightness $I'$ that has the form of an integer power $n \geq 1$ of the cosine of the zenith angle:
\begin{equation}
    I(r) = \cos^n\phi = (1 - r^2)^{n/2}.
\end{equation}
Inserting this into Eq. \ref{eq:hankel_solution} yields
\begin{equation}
    V_{{\rm cos}, n}(u) = \begin{cases} 
	 \frac{(u \pi )^{-\frac{n-1}{2}} \Gamma \left(\frac{n +3}{2}\right) J_{\frac{n+1}{2}}(2 \pi  u)}{n+1} & u > 0 \\
    \frac{2\pi}{n+1} & u=0
    \end{cases},
    \label{eq:gencosine}
\end{equation}
For $n=1$ this reduces to
\begin{equation}
V_{\rm cos}(u) \equiv V_{\rm cos, 1} = \frac{J_1(2 \pi u)}{u},
\label{eq:cossky}
\end{equation}
which we will refer to as the \textsc{cosza} pattern.

For $n=2$, we have
\begin{equation}
    V_{{\rm cos},2} = \frac{\sin(2\pi u) - 2\pi u \cos(2\pi u)}{4\pi^2 u^3}.
\end{equation}

For large values of $n$, this resembles a Gaussian bump centered on the zenith. The higher $n$, the narrower this bump. Varying the parameter $n$ thus allows one to test a transition between compact and diffuse structure.

\subsubsection{Polynomial Dome}
\label{sec:hankel:polydome}
A closely-related simple parametrized pattern takes the form
\begin{align}
    I'(r) &= 1 - r^{2n}, 
\label{eq:pattern_polydome}
\end{align}
which describes an inverted dome centred at zenith and falling to zero at the horizon ($r=1$).

We have
\begin{align}
    V_{{\rm pd},n}(u) &= V_{\rm cos}(u) - 2\pi \int_0^1 r^{2n+1}J_0(2\pi u r)\ dr.
\end{align}
This has the solution
\begin{align}  
    V_{{\rm pd},n}(u) = &V_{\rm cos}(u) \nonumber \\
    &- \begin{cases}
    \frac{\pi}{n+1} \pFq{1}{2}{n+1}{1, n+2}{-\pi^2 u^2}, & u > 0 \\
    \frac{2\pi n}{2n + 2} & u= 0
    \end{cases},
    \label{eq:polydome}
\end{align}
where ${}_p F_q$ is the generalized hypergeometric function defined in App.~\ref{app:special:hypergeometric}. We call the pattern \cref{eq:pattern_polydome} \model{polydome}.


In particular, for $n=1$ we find
\begin{equation}
    V_{\rm pd, 1} = V_{\rm cos}(u) - \frac{J_2(2\pi u) - \pi u J_3(2\pi u)}{\pi u^2},
    \label{eq:quaddome}
\end{equation}
which we call the \textsc{quaddome} solution.

\subsubsection{Projected Gaussian}
\label{sec:hankel:projgauss}
It is common to use Gaussians to describe components of resolved sources, since they are simple to Fourier transform. The simulator \code{PRISim}, for example, simulates diffuse emission by treating each pixel as a small Gaussian \citep{thyagarajan_nithyanandanprisim_2020}. It is of interest, therefore, to look for a solution for a Gaussian pattern. In this paper, we derive solutions for two forms of Gaussians, the first being the \emph{projected} Gaussian,
\begin{equation}
    I'(r) = e^{-r^2/\sigma^2},
\end{equation}
so-called because $I'$ includes the projection $1/\cos \phi$. The other form, solved in \S\ref{sec:convolutions:gaussian}, is also a Gaussian in $r$, but in the non-projected form.

Using our basic integral, \cref{eq:hankel_solution}, we have 
\begin{equation}
    V_{\rm pg}(u) = 2\pi \int_0^1 r \exp\left(\frac{-r^2}{\sigma^2}\right) J_0(2\pi u r) dr.
\end{equation}
We cannot integrate this directly, but we can find a converging series representation for the solution by Taylor-expanding the exponential or the Bessel function. Each expansion yields a different form for the solution, which admit different insights into the structure. Since the Taylor-series solutions can be evaluated to arbitrary precision independently of the properties of whatever simulated data they are compared against, they are still useful for simulator validation.

\paragraph{Large-$\sigma$ expansion.}
For $\sigma \gtrsim 1$, it makes sense to expand the exponential about $r=0$, because the exponential will have magnitude less than unity over the range of the integral. This yields
\begin{align}
    V_{\rm pg}(u) &= 2\pi \sum\limits_{k=0}^\infty \frac{(-1)^k}{k! \sigma^{2k}} \int_0^{1}dr\ r^{2k+1} J_0(2\pi u r) \nonumber  \\
    &= \pi \sum\limits_{k=0}^\infty \frac{(-1)^k}{\sigma^{2k}(1+k)!} \pFq{1}{2}{1+k}{1,2+k}{-\pi^2 u^2},
    \label{eq:projgauss:largesig}
\end{align}
where the second equality follows from \cref{eq:integral_of_power_j0} and properties of the hypergeometric function are given in App.~\ref{app:special:hypergeometric}. 
A useful outcome of this equation is that if $\sigma \rightarrow \infty$, only $k=0$ contributes at all, and we obtain $V_{\rm pg, \infty}(u) = J_1(2\pi u)/u$, as we expect from the solution for a cosine sky (cf. \cref{eq:cossky}).
App.~\ref{app:special:hypergeometric}, and particularly \cref{fig:hyp1f2_asymptotic}, demonstrate that the hypergeometric function is asymptotically almost constant, so the terms in the series are dominated by the prefactor $(-1)^k / (1+k)! \sigma^{2k}$. Thus, this expansion is absolutely convergent by the alternating series test. Nevertheless, we may only expect the series to decrease monotonically for $\sigma > 1$, yielding jinc-like\footnote{jinc: common usename for $J_0(x)/x$ for Bessel function of the first kind $J_0$.} solutions.
In practice, we find that the expansion works well down to $\sigma \sim 0.25$, and is preferred over the following small-$\sigma$ expansion for these values because it is stable for large $u$.

\paragraph{Small-$\sigma$ expansion.}
For $\sigma \lesssim 0.25$, the previous expansion is not efficient, and it is more useful to expand the Bessel function in its Taylor series.
For small-$\sigma$, the exponential cuts off the integrand for small $r$, rendering the argument to the Bessel function small over most of the important part of the integrand.

The Taylor expansion follows as:
\begin{align}
    V_{\rm pg}(u) &= 2\pi \sum\limits_{k=0}^\infty \frac{(-1)^k (\pi u)^{2k} }{(k!)^2}  \int_0^{1} r^{2k+1} \exp\left(-\frac{r^2}{\sigma^2}\right) \nonumber \\
    &= \sum\limits_{k=0}^\infty \frac{(-1)^k}{(k!)^2} \pi \sigma^2 (\pi u \sigma)^{2k} \left[\Gamma(k+1) - \Gamma(k+1, \frac{1}{\sigma^2})\right] \nonumber \\
    &= \pi \sigma^2 \left[e^{-\pi^2 \sigma^2 u^2} - \sum\limits_{k=0}^\infty \frac{(-1)^k}{(k!)^2} (\pi u \sigma)^{2k} \Gamma(k+1, \sigma^{-2})\right],
    \label{eq:projgauss:smallsig}
\end{align}
where $\Gamma(a,b)$ is the upper-incomplete Gamma function defined in \cref{eq:inc_gamma}. The last equality uses the Taylor series of an an exponential function.

Here, for small $\sigma$, the second term becomes less and less influential, so that 
\begin{equation}
    V_{\rm pg}(u|\sigma \ll 1) = \pi \sigma^2 e^{-\pi^2 \sigma^2 u^2},
\end{equation}
which is exactly the Fourier transform of $I'$, as expected.\footnote{In this function and elsewhere, we use a vertical bar to denote a condition of the function. In this case, the notation indicates this is the solution for $\sigma \ll 1$.} This solution thus characterizes deviations from this limit by the term containing the sum. 


Note that the small-$\sigma$ expansion exhibits poor convergence for large $u$. 
However, at large $u$, the Bessel function may be sufficiently damped and regular to render the remainder of the integral to infinity negligible. The integral to infinity is simply the standard Gaussian solution. Thus we expect that deviations from the Gaussian solution decline at high $u$. 
In practice we advocate using \cref{eq:projgauss:smallsig} for increasing values of $u$, until the solutions are within a desired tolerance of the purely Gaussian term, and from then, for larger $u$, using the Gaussian directly.


\subsection{Convolution Method}
\label{sec:convolutions}

We'll now move on to the second approach for finding patterns and solutions to the RIME. Instead of writing the horizon cutoff as an integration limit, we may instead write it as a step function $W$ multiplied by the brightness $I$. This allows us to extend the integral over an ``infinite'' $(l,m)$-plane, and consequently turns the integral into a standard 2D Fourier transform in $(l,m)$ to Fourier duals $(u, v)$.

Letting $I \rightarrow WI$, with
\begin{equation}
    W(l,m) = \begin{cases}
       1 & \sqrt{l^2 + m^2} \leq 1, \\
       0 & \sqrt{l^2 + m^2} > 1,
       \end{cases}
	\label{eq:horizon_window}
\end{equation}

we may write
\begin{equation}
    V(u,v,w) = \int\limits_{-\infty}^\infty \int\limits_{-\infty}^\infty \frac{\ W(l,m) }{n} I(\mathbf{l}) e^{-2\pi i [\mathbf{u}\cdot\mathbf{l} - w n)]} \: {\rm d}l {\rm d}m.
    \label{eq:useable_visibility}
\end{equation}

This is the Fourier transform of a product of two functions: the window $W\exp(-2 \pi i w n)/n$ and the beam-weighted sky surface brightness $I$. 
By the convolution theorem, the Fourier transform of this product is the convolution of the respective Fourier transforms of each term:
\begin{equation}
    V(\vecb{u}) = \tilde{W}(\vecb{u}) \ast \tilde{I}(\vecb{u}),
	\label{eq:conv_full}
\end{equation}

Evaluating the window function transform, we may invoke circular symmetry and apply the Hankel transform (in the following, $|\vecb{u}| = \sqrt{u^2 + v^2}$):
\begin{align}
    \widetilde{W} &= \int \frac{W(l,m)}{\sqrt{1 - l^2 - m^2}} e^{2\pi i w \sqrt{1 - l^2 - m^2}} e^{-2 \pi i (ul + vm)} dl dm \nonumber \\
    &= 2 \pi \int\limits_0^\infty \frac{r W(r) e^{2\pi i w \sqrt{1 - r^2}}}{\sqrt{1-r^2}} J_0(2 \pi |\vecb{u}| r) dr   \nonumber \\
    &= 2 \pi \int\limits_0^1 \frac{r e^{2\pi i w \sqrt{1 - r^2}}}{\sqrt{1-r^2}} J_0(2 \pi |\vecb{u}| r) dr \nonumber  \\
    &= 2 \pi \sum_{n=0}^\infty \frac{(2\pi i w)^n}{n!} \int\limits_0^1 r (1-r^2)^{\frac{n-1}{2}} J_0(2 \pi |\vecb{u}| r) dr \nonumber \\
    &= 2 \pi \sum_{n=0}^\infty \frac{(2\pi i w)^n}{n!(n+1)} {}_0F_1\left(\frac{n+3}{2}, -\pi^2 |\vecb{u}|^2\right) \nonumber \\
    &= 2 \pi \sum_{n=0}^\infty (2\pi i w)^n \frac{\Gamma\left(\frac{n+3}{2}\right)}{\Gamma(n+2)} J_{\frac{n+1}{2}}(2\pi |\vecb{u}|) (\pi |\vecb{u}|)^{-\frac{n+1}{2}}
    \label{eq:w-expansion}
\end{align}

When $w = 0$ (the fiducial case), all of the summands are zero except for $n=0$, giving
\begin{align}
    \widetilde{W}(|\vecb{u}|) &= 2\pi {}_0F_1(3/2, -\pi^2 |\vecb{u}|^2) \nonumber \\
    &= 2 \pi\: \sinc(2 \pi |\vecb{u}| ).
	\label{eq:window_w0}
\end{align}
For this case, we're left with the following expression for the visibility:
\begin{equation}
    V(\vecb{u}) = 2\pi {\rm sinc}(2\pi u) * \tilde{I}(\vecb{u}),
\end{equation}
where $I$ and $\tilde{I}$ are defined over the entire $(l,m)$ and $(u,v)$ planes, respectively. This has the advantage that in principle, $I$ is not constrained to be symmetric around zenith. 
It also permits some natural closed-form solutions that the direct integration method did not.

If we \textit{do} assume that $I$ is symmetric about zenith, we can convert this expression to polar coordinates,
\begin{equation}
    V(u) = \int_0^\infty du'\int_0^{2\pi} d\theta\  \sin(2\pi u') \tilde{I}(\sqrt{u^2 + u'^2 - 2uu'\cos\theta}),
    \label{eq:conv:axisymmetric}
\end{equation}
taking (without loss of generality) $\theta$ to be the angle between $\vecb{u}$ and $\vecb{u}'$. Note that the $2 \pi {\rm sinc}(2\pi u) = \sin(2\pi u)/ u$, and the $1/u$ cancels with the factor of $u$ in the integration measure $d^2 u = u\:{\rm d}\theta {\rm du}$ in the convolutional integral.

As an aside, we note that the derivation of $V(\vecb{u}) = \tilde{W} \ast \tilde{I}$ may offer a route to faster simulations. Wide, diffuse power on the sky tends to be fairly compact in $(u,v)$ space, so that the convolution of $\widetilde{W}$ with $\tilde{I}$ may be done approximately over fewer $(u,v)$ pixels than would be required for integrating over the sky. This is similar to how the \emph{Fast Holographic Deconvolution} (FHD) package takes advantage of the relative compactness of wide beams in $(u,v)$ space to expedite forward modeling \citep{sullivan2012}.

\subsubsection{Monopole}
\label{sec:convolutions:monopole}
The simplest possible diffuse pattern is a monopole:
\begin{equation}
    I(r) = 1,  
\end{equation}
which has the Fourier transform $\tilde{I}(\vecb{u}) = \delta(\vecb{u})$.
Using \cref{eq:conv_full} directly results in 
\begin{equation}
    \label{eq:monopole}
    V_{\rm mono}(u) = \widetilde{W}(\vecb{u}),
\end{equation}
where $\widetilde{W}$ is either the series of \cref{eq:w-expansion} or \cref{eq:window_w0} for $w=0$. This gives us one pattern which can be used to test a simulator's capability to handle baselines with $w\neq 0$.






\subsubsection{Gaussian}
\label{sec:convolutions:gaussian}
We will now solve the RIME for a non-projected Gaussian, as mentioned in \S\ref{sec:hankel:projgauss}. This non-projected Gaussian in $r = \sqrt{l^2 + m^2}$ is a traditional basis for describing extended structures and is typically used as a restoring beam during image reconstruction. First we consider a Gaussian centered at $\vecb{l} = 0$ (zenith) and later move to an arbitrary location.

Start with
\begin{equation}
    I_{\rm g}(r) = e^{-l^2/2\sigma^2},
 \label{eq:conv:gaussmodel}
\end{equation}
such that 
\begin{equation}
    \tilde{I}_{\rm g}(u) = a^2 e^{-\pi a^2 u^2}, \ \ \ a^2 = 2\pi\sigma^2,
\end{equation}
Inserting this into \cref{eq:conv:axisymmetric} and integrating over $\theta$ yields
\begin{equation}
    \label{eq:gaussian}
    V_{\rm g}(u) = 2 \pi \tilde{I}_{\rm g}(u) \int_0^\infty du'\ \sin(2\pi u') e^{-\pi a^2 u'^2} I_0(2\pi u u' a^2),
\end{equation}
where $I_0$ is the zeroth-order modified Bessel function of the first kind.

The remaining integral does not have a known closed-form solution, but we can look for series solutions as in \S\ref{sec:hankel:projgauss} by considering the small- and large-$a$ regimes in turn.



\paragraph{Expansion for large $a$}
For large $a$ (relative to $u$), the integrand is dominated by the exponential and falls off steeply for small $u'$. 
In this case, we may expect the sine term to be well-approximated by a low-order Taylor expansion about zero, as the bulk of the integrand's density is at low $u'$. 
Using the Taylor expansion of sine,
\begin{equation}
    \sin(x) = \sum_{k=0}^\infty (-1)^k\frac{x^{2k+1}}{\Gamma[2(k+1)]},
\end{equation}
we have 
\begin{align}
    \begin{split}
    V_{\rm g}(u) ={}& 2\pi \tilde{I}_{\rm g}(u)  \sum_{k=0}^\infty \frac{(-1)^k(2\pi)^{2k+1}}{\Gamma(2(k+1))} \\
     &\times\int_0^\infty du'\ u'^{2k+1}  e^{-\pi a^2 u'^2} I_0(2\pi u u' a^2) \end{split} \nonumber\\
    ={}& \pi^{3/2} \sum_{k=0}^\infty \left(\frac{-\pi}{a^2}\right)^{k}\frac{L_k(-\pi a^2u^2)}{\Gamma(k + 3/2)},
    \label{eq:gauss_large_a_laguerre}
\end{align}
where $L_k$ is the $k^{\rm th}$-order Laguerre polynomial defined in \cref{eq:laguerre_def}.
For $a\rightarrow\infty$, the Laguerre polynomial is dominated by its highest-order term, such that $L_n(x\rightarrow\infty) \approx (-1)^n x^n / n!$.
In this case, the visibility reduces to
\begin{align}
    V_{\rm g}(u|a\rightarrow\infty) &= \pi^{3/2} \sum_{k=0}^\infty \frac{(-\pi^2 u^2)^k}{\Gamma(3/2+k)\Gamma(1+k)} \nonumber \\
    &= 2\pi \sinc(2\pi u),
\end{align}
which is expected, since in that limit the Gaussian approaches a monopole.

While this expansion is simple to evaluate, it only converges quickly for large $a$ and small $u$ --- for $u \gtrsim 1/\pi$, the series initially increases before the $\Gamma$ terms begin to dominate and bring the series to convergence.

Writing the Laguerre polynomial in its power-series expansion (cf. \cref{eq:laguerre_def}) we have
\begin{align}
    V_{\rm g}(u) &= \pi^{3/2} \sum_{k=0}^\infty \sum_{j=0}^k \left(\frac{-\pi}{a^2}\right)^{k}\frac{k! (\pi a^2u^2)^j}{(j!)^2(k - j)!\Gamma(3/2+k)} \nonumber \\
    &= \pi^{3/2} \sum_{j=0}^\infty \sum_{k=j}^\infty \left(\frac{-\pi}{a^2}\right)^{k}\frac{k!(\pi a^2u^2)^j}{(j!)^2(k - j)!\Gamma(3/2+k)} \nonumber \\
    &= \pi^{3/2} \sum_{j=0}^\infty (-1)^j (\pi u)^{2j} \frac{\pFq{1}{1}{1+j}{3/2+j}{-\pi / a^2}}{j!\Gamma(3/2+j)}.
    \label{eq:gaussian_large_a_1f1}
\end{align}
where in the second equality we merely swapped the summands.
This expansion swaps the conditions for convergence -- it is rapidly converging for small values of the product $\sim au^2$. That is, it requires \textit{very small} values of $a$ for most reasonable values of $u$ to be numerically well-behaved. To obtain an expression with faster convergence, we expand the hypergeometric function as a sum over $k$ as per its definition (cf. \cref{eq:hypergeometric_def}) and then perform the sum over $j$. The result is
\begin{equation}
    V_{\rm g}(u) = \pi^{3/2} \sum_{k=0}^\infty \left(\frac{-\pi}{a^2}\right)^k \, \frac{\pFq{1}{2}{k+1}{1,k+3/2}{-\pi^2 u^2}}{\Gamma(k+3/2)}. 
    \label{eq:gaussian_sine_expansion}
\end{equation}
Note the similarity of this equation to \cref{eq:gauss_large_a_laguerre}, with the Laguerre polynomial being replaced by the hypergeometric function. 
Interestingly, the Laguerre polynomial is not equivalent to the ${}_1F_2$ function in this form --- the Laguerre polynomial depends on $a$, while the hypergeometric function does not. Nevertheless, the result of an infinite series is the same. The essential trick in obtaining \cref{eq:gaussian_sine_expansion} from \cref{eq:gauss_large_a_laguerre} seems to be the swapping of the summands.

\begin{figure}
    \centering
    \includegraphics[width=\linewidth]{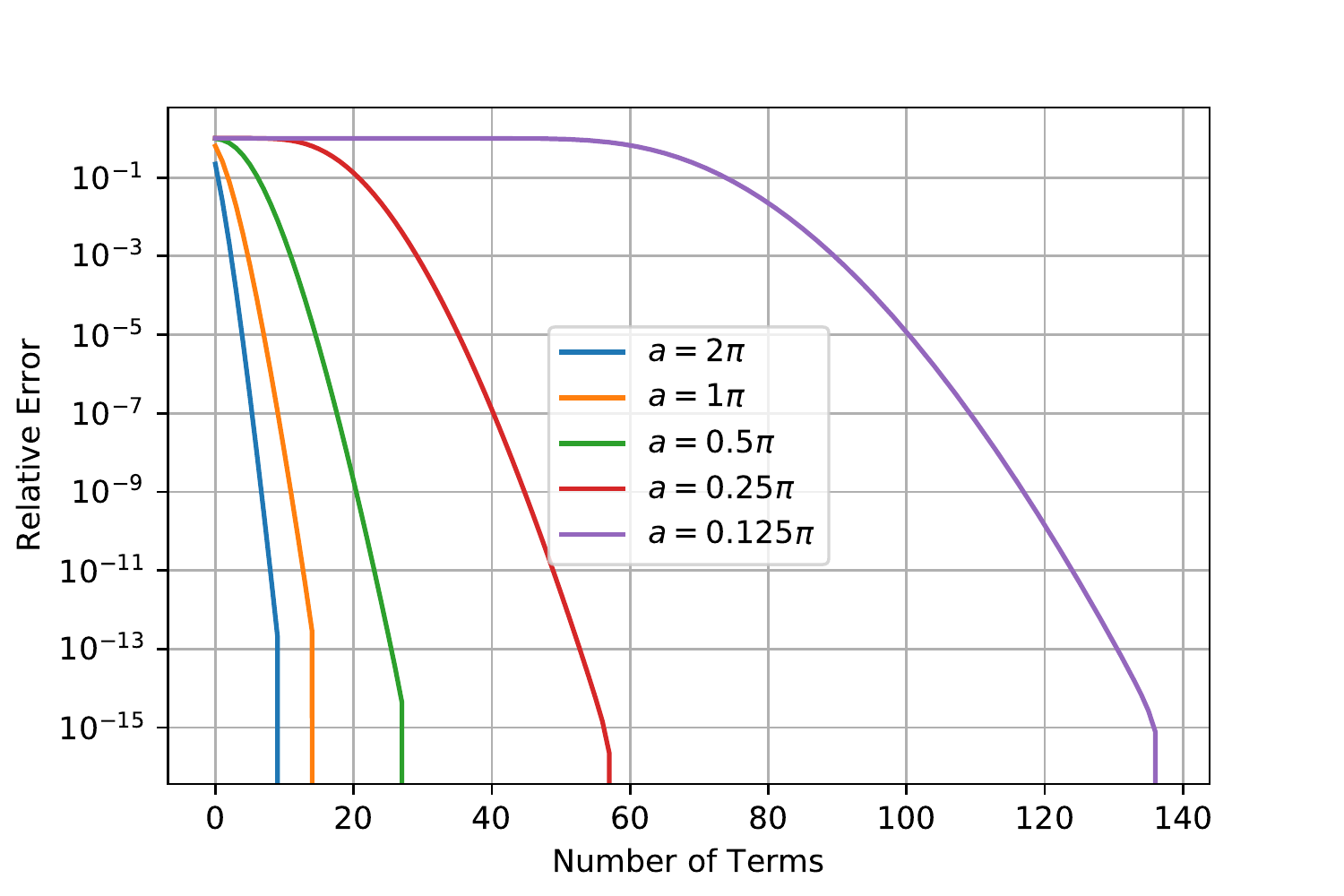}
    \caption{Relative error of the cumulative sum from \cref{eq:gaussian_sine_expansion} as a function of the number of terms used. All curves use the large value of $u=200.3$, thus showing a conservative case. The expansion converges to machine precision in $\lesssim 100$ terms for $a \gtrsim \pi/8$.}
    \label{fig:gaussian_sine_expansion}
\end{figure}

The usefulness of this particular expansion is that, regardless of the value of $u$, the series is dominated by the power-law and $\Gamma(k+3/2)$ (see justification in App.~\ref{app:special:hypergeometric}), which is absolutely convergent. 
Despite being theoretically convergent, for small $a$ the series will not be numerically stable (rising to extremely large values before falling again).
In \cref{fig:gaussian_sine_expansion} we show the behaviour of the expansion, and note that it converges in a reasonable number of terms for $a \gtrsim \pi/8$, even for large $u$.

\paragraph{Expansion for small $a$}
We now look to the case of small $a$. Although \cref{eq:gaussian_large_a_1f1} is convergent for \emph{very} small $a$ (or small $u$), it would be useful to derive an expansion that works for ``reasonably small'' $a$ (say, $a \lesssim 0.1$) at all typical $u$. 
When $a$ and $u$ are small, the integral is dominated by small arguments of both the exponential and $I_0$, and it makes sense to expand either of these in its power series. It is simpler to expand the Bessel function (cf. \cref{eq:besselI_taylor}):
\begin{align}
        \frac{V_{\rm g}(u)}{\tilde{I}_{\rm g}(u)} &=  2\pi \sum_{k=0}^\infty \frac{(\pi u a^2)^{2k}}{\Gamma^2(k+1)}
        \int \limits_0^\infty  \sin(2\pi u') e^{-\pi a^2 u'^2} u'^{2k} \,du' \nonumber \\
    &= 2\pi \sum_{k=0}^\infty \frac{(\pi a^2 u^2)^{k}}{a^2 k!} \pFq{1}{1}{1+k}{3/2}{-\pi/a^2}.
\label{eq:gaussian_expansion}
\end{align}
This is the defining function of the small-$a$ expansion. 
In the limit $a\rightarrow0$, only $k=0$ contributes, which leaves
\begin{equation}
    \frac{V_{\rm g}(u|a\rightarrow0)}{\tilde{I}_{\rm g}(u)} = 2\pi e^{- \pi/a^2} \frac{ {\rm erfi}(\sqrt{\pi}/a)}{2 / a} 
    = 1,
\end{equation}
in which the second equality uses the zeroth-order asymptotic expansion of ${\rm erfi}(x)$. Thus we obtain the Gaussian solution we expect in the flat-sky / narrow-field approximation, in which the RIME reduces to a 2D Fourier transform.

\begin{figure}
    \centering
    \includegraphics{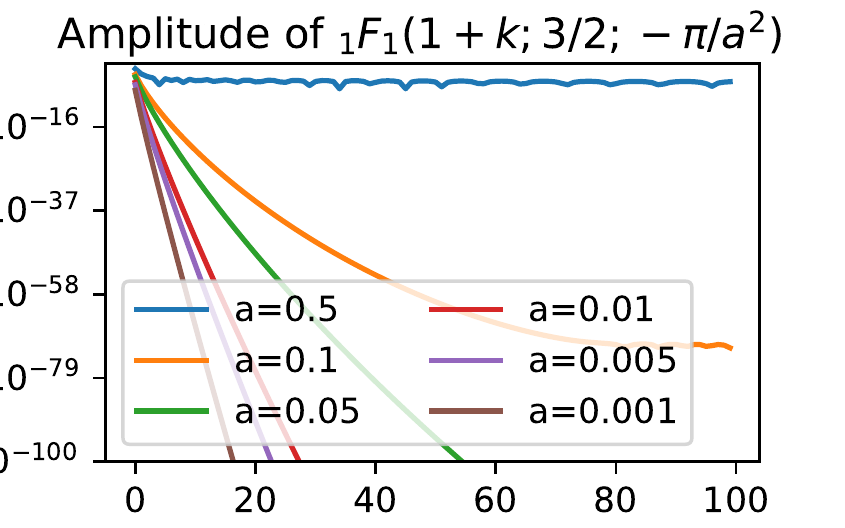}
    \caption{The series $\pFq{1}{1}{1+k}{3/2}{-\pi/a^2}$ for several reasonable values of $a$. It shows rapid decrease for small $a$.}
    \label{fig:hyp1f1_convergence}
\end{figure}

\begin{figure*}
    \centering
    \includegraphics[width=0.9\linewidth]{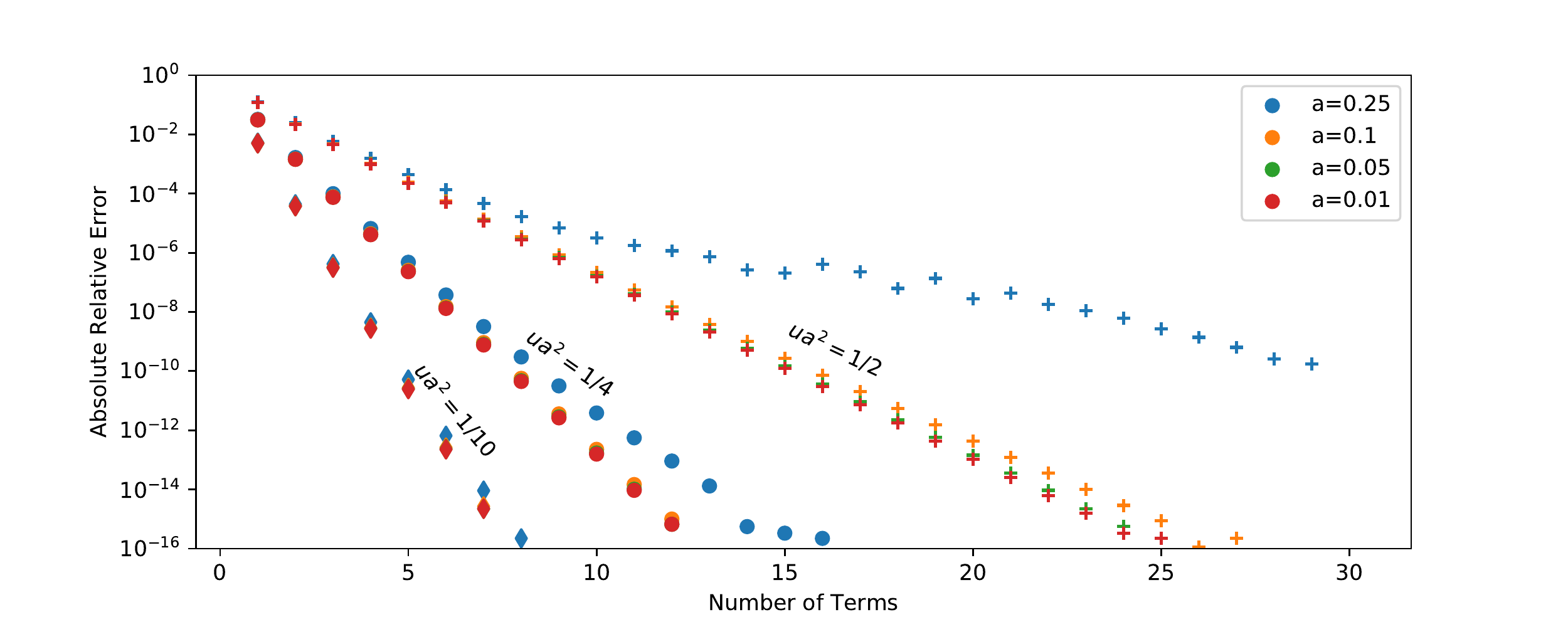}
\caption{Convergence of \cref{eq:gaussian_expansion} for constant values of $u a^2$. Note the strong dependence on this quantity, and weak dependence on $a$.}
    \label{fig:gaussian_small_a_convergence}
\end{figure*}

Figure \ref{fig:hyp1f1_convergence} shows the behaviour of the ${}_1F_1$ factor, which decreases rapidly with $k$ for $a < 1/3$. This means that the full series is absolutely convergent. From trial and error, it appears that the rate of convergence is strongly dependent on the value of $ua^2$, with marginal dependence on the absolute values of $a$ and $u$ individually. See \cref{fig:gaussian_small_a_convergence} for a summary of this convergence rate. We find that for $ ua^2 \lesssim 2/3$, less than 30 terms are required for double-precision accuracy, as long as $a \lesssim 0.25$. However, we also note that the evaluation of ${}_1F_1$ is slower for larger $a$.
For larger values of $ua^2$, the series may become numerically unstable.

Note that the dependence on $u a^2$ here means that larger $u$ may converge for the same small $a$ as compared to \cref{eq:gaussian_large_a_1f1}, which makes it ideal for the small $a$ considered in this section.

Further convergence properties and approximations to \cref{eq:gaussian_expansion},  can be found in App.~\ref{app:gaussian_smalla}. Indeed, for $a \lesssim 0.1$, the approximations in App.~\ref{app:gaussian_smalla} are extremely useful in practical situations, and we encourage their usage in place of \cref{eq:gaussian_expansion}. 
The appendix also includes a first-order estimate of the error induced by making the common assumption that the visibility function of a Gaussian source is a Gaussian in $\vecb{u}$. This gives some idea of the range of validity of the flat-sky approximation for a Gaussian source. We show that such an approximation leads to errors greater than 1\% for Gaussians that a given baseline can ``barely resolve'', for baselines shorter than $\sim 1700 \lambda$. Thus, the more accurate expansions of this section are required for the extreme accuracy desired by current low-frequency arrays.

\subsubsection{Non-centred Gaussian}

So far all the derived patterns have been axisymmetric about the zenith phase center. Fortunately, the convolutional approach offers a way to find patterns which are not symmetric about the zenith by shifting the center of an axisymmetric pattern from the zenith and applying the Fourier shift theorem to an existing solution. As an example of this, consider a pattern $I$ which is axisymmetric about the zenith, and from it define a \emph{shifted} pattern $I_S$ which is axisymmetric around some point $\vecb{l}_0$, i.e. $I_{S} = I(\vecb{l} - \vecb{l}_0)$.
By the Fourier shift theorem we have that
\begin{equation}
    \tilde{I}_{S}(\vec{u}) = e^{-2\pi i \vecb{l}_0 \cdot \vecb{u}} \tilde{I}(|u|),
\end{equation}
which means that
\begin{multline}
    V_{S}(\vecb{u}) = e^{-2\pi i \vecb{l}_0 \cdot \vecb{u}} \int_0^{2\pi} d\theta \int_0^\infty du'\,\sin(2\pi u') \\
      \times e^{2\pi u' i (l_0^x \cos \theta + l_0^y \sin\theta)} \tilde{I}(|\vecb{u} - \vecb{u}'|),
	\label{eqn:convolutional:offzen}
\end{multline}
where again we have taken $\theta$ (without loss of generality) to be the angle between $\vecb{u}$ and $\vecb{u}'$. 
Note that there is now an extra degree of freedom $\gamma$ in the relative orientation of $\vecb{l}_0$ with respect to $\vecb{u}$, so that $l_0^x = |l_0| \cos\gamma$.
This is applicable to any pattern that is axisymmetric about a given off-zenith direction. 

Letting $I = \exp\left[ -l^2 / (2 \sigma^2)\right]$ as in \cref{eq:conv:gaussmodel}, and noting that 
\begin{equation}
    \int_0^{2\pi} d\theta e^{a \cos\theta + b \sin\theta} = 2\pi I_0(\sqrt{a^2 + b^2}),
\end{equation}
we find
\begin{multline}
    \frac{V_{SG}(\mathbf{u})}{\tilde{I}_{\rm g}(u)} = 2\pi  e^{-2\pi i \vecb{u}\cdot \vecb{l}_0}  \int_0^\infty du'\ \sin(2\pi u') e^{-\pi a^2 u'^2} \\
    \times I_0\left(2\pi u' \sqrt{u^2a^4 - l_0^2 + 2 i a^2 \vecb{u}\cdot\vecb{l}_0}\right),
\end{multline}
Letting $v = \sqrt{u^2 - l_0^2/a^4 + 2 i \vecb{u}\cdot\vecb{l}_0/a^2}$, the integrand has the same form as in \cref{eq:gaussian}:
\begin{align}
    \label{eq:shifted_gaussian}
    \frac{V_{SG}(\mathbf{u})}{\tilde{I}_{\rm g}(u)} = 2\pi 
    e^{-2\pi i \vecb{u}\cdot \vecb{l}_0} \int_0^\infty du'\, &\sin(2\pi u') e^{-\pi a^2 u'^2} \nonumber \\
    &\times I_0\left(2\pi u' v a^2 \right).
\end{align}
As in \S\ref{sec:convolutions:gaussian}, we can either expand the sine when the Bessel function dominates the shape of the integrand, or expand the Bessel function. Unlike before, however, the Bessel function will dominate for large $l_0$ as well as for large $a$. Thus, for the limit of large $a$ and/or $l_0$, characterized by $|v| a^2 > 1$, we expand the Bessel function as in  \cref{eq:gauss_large_a_laguerre}. Notably, the integral over $u'$ does \emph{not} give us an exponential that cancels with the Gaussian exactly:
\begin{align}
    \begin{split}
    V_{SG}(\mathbf{u}) ={}& 2\pi \tilde{I}_{\rm g}(u) 
    e^{-2\pi i \vecb{u}\cdot \vecb{l}_0}
    \\
    &\times\sum_{k=0}^\infty \frac{(-1)^k(2\pi)^{2k+1} k!}{2(2k+1)! (\pi a^2)^{k+1}} e^{\pi a^2 v^2} L_k(-\pi a^2v^2) \end{split} \nonumber \\
    \begin{split}
    ={}& \pi^{3/2} e^{-2\pi i \vecb{u}\cdot \vecb{l}_0} e^{-\pi a^2(u^2 - v^2)} \\
    &\times \sum_{k=0}^\infty \left(\frac{-\pi}{a^2}\right)^{k}\frac{L_k(-\pi a^2v^2)}{\Gamma(3/2+k)},
    \end{split}
\end{align}
This is the same result as \cref{eq:gauss_large_a_laguerre} (and admits a similar expansion as \cref{eq:gaussian_sine_expansion} as well), except in terms of $v$ and with an extra phase term in front. The phase term can be further simplified by expanding $v$ in the exponential:
\begin{equation}
    e^{-2\pi i \vecb{l}_0 \cdot \vecb{u} } e^{-\pi a^2 (u^2 - v^2)}
    = e^{-\pi l_0^2/a^2},
\end{equation}
Altogether, this gives us the forms
\begin{align}
    V_{SG}(\mathbf{u}) &= \pi^{3/2} e^{-\pi l_0^2/a^2}  \sum_{k=0}^\infty \left(\frac{-\pi}{a^2}\right)^{k}\frac{L_k(-\pi a^2v^2)}{\Gamma(k+3/2)} \\
    &= \pi^{3/2} e^{-\pi l_0^2/a^2} \sum_{k=0}^\infty \left(\frac{-\pi}{a^2}\right)^k \, 
    \frac{\pFq{1}{2}{1+k}{1,3/2+k}{-\pi^2 v^2}
    }{\Gamma(k+3/2)}
	\label{eqn:dispgaussian_largea}
\end{align}
for large-$a$ Gaussians. For the same reasons as explained in \S\ref{sec:convolutions:gaussian}, this converges absolutely, with numerical stability for $a\gtrsim \pi/8$ independent of $u$ and zenith offset $l_0$.

For $l_0 = 0$, $v \rightarrow u$ and we recover the on-zenith solution. Also, for $a \rightarrow \infty$, the preceding exponential cancels to unity and again, $v \rightarrow u$, so we recover the perfectly real monopole solution \cref{eq:monopole}. This is consistent with the expectation that if the Gaussian is sufficiently large compared with its offset from zenith, the offset becomes negligible.


For small $a$, we expand $I_0$ in a power series as in the previous section: 
\begin{align}
    \frac{V_{\rm SG}(\vecb{u})}{\tilde{I}_{\rm g}(u)} &= 2\pi e^{-2\pi i \vecb{u}\cdot \vecb{l}_0} \sum_{k=0}^\infty \frac{(\pi a^2 v^2)^{k}}{a^2 k!} 
    \pFq{1}{1}{1+k}{3/2}{-\frac{\pi}{a^2}},
	\label{eqn:dispgaussian_smalla}
\end{align}
Recall that while being absolutely convergent in theory, the convergence rate of the on-zenith Gaussian solution depended strongly on  $ua^2$ (the smaller the faster the convergence). For this off-zenith solution, this translates into $|v|a^2$:
\begin{align}
|v| a^2 = a^2 \sqrt{
\left( u^2 - \frac{l_0^2}{a^4}  \right)^2 + \left( \frac{2 \vecb{l}_0 \cdot \vecb{u}}{a^2} \right)^2} \leq \sqrt{l_0^2 + a^4 u^2}
\label{eq:va2inequality}
\end{align}
where the inequality comes from $\vecb{l}_0 \cdot \vecb{u} \leq l_0 u$. 
Noting our previous result that the series converges well for $|v|a^2 < 2/3$, it thus may be difficult to reach for any $u$ or $a$ near the horizon, where $l_0 \rightarrow 1$.

The shifted Gaussian presented in this section represent arbitrary Gaussian sources, from which generalized diffuse sky models may be built up by linear combination, thus the solutions are of quite general usefulness.
Nevertheless, we must caveat the usefulness to some degree; for wide gaussian sources, we have a solution (Eq. \ref{eqn:dispgaussian_largea}) that converges well for any baseline length or source position. However, for highly compact sources, we have a solution (Eq.~\ref{eqn:dispgaussian_smalla}) that converges well only for limited baseline length (up to $u \sim 1/a^2$) and for sources not too close to the horizon (cf. Eq.~\ref{eq:va2inequality}). Thus there are regions of parameter space for which none of these solutions apply well (numerically speaking). Deriving such solutions is the subject of future work.

\subsection{Separation of Variables}
\label{sec:separable}

The solutions in the previous subsections all feature symmetry about some axis. Solutions derived with the Hankel transform approach require rotational symmetry about the zenith, due to the inclusion of the $\sqrt{1 - l^2 - m^2}$ factor in defining $I'$. The patterns found in \S\ref{sec:convolutions} are also rotationally symmetric, either about the zenith or about some point off-zenith.\footnote{The off-zenith shift for convolutional solutions \emph{does} break rotational symmetry, because the horizon is no longer symmetric about the center of the function. However, we may be interested in patterns that have an angular dependence other than at the horizon.} While the convolutional approach doesn't inherently require such symmetry, all of the convolutional solutions derived here feature such symmetry because it reduces the convolution integral to a simpler 1D form.

One way to introduce asymmetry is to make the integrand of the RIME separable into two independent coordinates and use independent 1D patterns for each. To achieve this, we define $x$ and $y$ as orthogonal axes rotated some angle $\xi$ from the $l$ and $m$ axes, such that $\vecb{x} = (x,y)$ is related to $\vecb{l}$ via $\vecb{x} = R_\xi \vecb{l}$ for rotation matrix $R_\xi$. We then define a the brightness function to have the form,
\begin{equation}
    I(x,y) = 
    \begin{cases}
        h(x) g(y) \sqrt{1 - x^2 - y^2} & |x| \text{ and } |y| < \frac{1}{\sqrt{2}}\\
        0 & \text{otherwise},
    \end{cases}
    \label{eq:separable_I}
\end{equation}
for some functions $h$ and $g$. The limits on $|x|$ and $|y|$ describe a square region inscribed within the horizon, which lets us carry the integral to the horizon while maintaining separability of the function. Inserting this into \cref{eq:vis_lmn} with $w=0$, we obtain
\begin{align}
    V_{\rm sep}(\vecb{u}) &= \int h(x) g(y) \sqrt{1 - x^2 - y^2} e^{-2 \pi i \vecb{l} \cdot \vecb{u}} \frac{d^2 l}{\sqrt{1 - l^2  - m^2}} \nonumber \\
    &=\int h(x) g(y) e^{-2 \pi i \vecb{l} \cdot (R_\xi^{-1} \vecb{u})} dx dy\nonumber \\
    &= \int\limits_{-1/\sqrt{2}}^{1/\sqrt{2}} h(x) e^{-2 \pi i x u'_x} dx
       \int\limits_{-1/\sqrt{2}}^{1/\sqrt{2}} g(y) e^{-2 \pi i y u'_y} dy
       \label{eq:separated_sincs}
\end{align}
In the last step, we've defined $\vecb{u}' = (u'_x, u'_y) \equiv R^{-1}_\xi \vecb{u}$, and switched the integration variables. This is solvable for many choices of $h$ and $g$.

As an example, we will let $h(x) = \sinc(a x)$ and $g(y) = \sinc(a y)$, where $a$ is a constant. This pattern is able to capture -- along with effects of baseline orientation -- various size-scale oscillations on the sky placed at different elevations.
In this case, each integral has a closed form solution in terms of sine integrals (cf. \S\ref{app:special:sine}). The solution is
\begin{equation}
    V_{\rm sinc}(\vecb{u}) = V_{\rm sinc}(u'_x)V_{\rm sinc}(u'_y)
\end{equation}    
with
\begin{equation}
    V_{\rm sinc}(u') = \frac{\text{Si}[(a + 2 \pi  u')/\sqrt{2}]+\text{Si}[(a - 2 \pi  u')/\sqrt{2}]}{a}
    \label{eq:xysinc}.
\end{equation}

We will refer to this pattern as \model{xysincs}. Its solution has roughly the form of a square rotated by an angle $\xi$ in the UV plane with sides of length $a/\pi$. If the integrals could be taken to $\pm \infty$, each would become a boxcar function.

This pattern is useful for a few reasons. It has two tunable parameters -- $a$ and $\xi$. Adjusting $a$ sets a scale for fluctuations on the sky, while changing $\xi$ changes the orientation of the pattern. Separate values of the constant $a$ may be chosen for $x$ and $y$, changing the shape to be more rectangular than square, but we will keep a single value of $a$ in this paper. Further, the solution does not smoothly taper off with increasing baseline length as the other solutions do --- Along each axis of the box, the solution remains mostly constant up to $|u| = a/\pi$, after which it drops off rapidly. This allows one to test for situations when one would expect a strong response on a longer baseline. Also, this pattern has a sharp dependence on baseline angle, which may reveal the effects of subtle baseline alignment errors.

This concludes our list of test patterns and their corresponding visibility functions. In the following sections, we will describe a basic visibility simulator and demonstrate the usefulness of these patterns for testing its integration scheme.

\section{An Integrator to Validate}
\label{sec:simulator}

The basic task of a simulator is to forecast instrument output within some desired accuracy. Broadly speaking, inaccuracy can come from missing information about the signal chain, errors in the instrument or sky model, or from errors introduced in the numerical integration process. It is this last type of error that the test patterns in the previous section are designed to handle. In this section we use our solutions to test the accuracy of a numerical integrator of our own devising.

A simulator of diffuse emission must numerically evaluate the integral in the RIME, whether by choosing a set of nodal points on the sky and integrating by quadrature, or transforming to a more compact harmonic space. If performing a quadrature sum, the contribution of the fringe term is approximated by its value at those nodal points. If working in harmonic space (e.g. the m-mode formalism of \cite{shaw2014}), the fringe term may be treated exactly but the discrete nature of the map means that the multipole expansion of the sky model will be approximate.

It is common for full-sky maps of diffuse emission to be published as \code{healpix} maps \citep{gorski_healpix_2005}. Any published map will have some finite resolution, whether that be spatial (pixels) or Fourier (multipoles). For pixelized maps, it is intuitive to choose the pixel centers as quadrature nodes for numerical integration, allowing the exact pixel values to be used in the quadrature sum. However, sky models may usually be interpolated  to scales smaller than this resolution limit to better evaluate the integral. 

Numerical integration over spherical surfaces are an industry unto themselves \citep[e.g.][]{Hesse2015,atkinson_1982}, and a full exploration of different integration schemes and their pros and cons is left to future work. For the purpose of demonstrating the usefulness of the patterns derived in \S\ref{sec:analytic_solns}, we choose as our integration scheme a simple 2D Riemann sum over \code{healpix} maps. This treats each map pixel with brightness temperature $T_p$ as a point source with flux $I_p = \Omega T_p$, where $\Omega$ is the pixel solid angle (which is the same for all pixels in \code{healpix}):
\begin{equation}
    V(\vecb{u}) = \sum\limits_{p\in I} I_p e^{-2 \pi i\hat{s}_p \cdot \vecb{u}}.
    \label{eq:rime_sim}
\end{equation}
We refer to this as the \emph{point source approximation} (PSA). This is implemented in the simulation package \code{healvis}.\footnote{\url{https://github.com/rasg-affiliates/healvis}} With \code{healvis}, we simulate visibilities from the test patterns and compare these to separate evaluations of the corresponding solutions.

The ``pixels = point sources'' integration method carries the potential for several errors. First, because we are modeling quadralateral healpix pixels as unresolved points the sampling is not continuous and will break down at resolutions approaching the pixel size. This error can be directly estimated by calculating how fringes vary across the \code{healpix} pixels. Such an error estimate will be a useful comparison point when evaluating our test patterns.

\subsection{An Error Bound for the PSA}
\label{sec:errorbound}

While upper-bounds for integration errors are a very active topic, relatively few methods relating to 2D integration over the sphere for an arbitrary set of chosen locations are to be found in the literature.
We use a result from \citet{Hesse2015}, citing a result of \cite{Freeden1998}, for functions that are Lipschitz-continuous: 
\begin{equation}
    |I_{\rm true} - I_{\rm num}| \leq 2\pi \sigma\  \underset{1\leq j \leq N_{\rm pix}}{\rm max} \left\{ {\rm diam}(P_j)\right\},
\end{equation}
where $I_{\rm true}$ and $I_{\rm num}$ are the true value of the integral and the numerical approximation, $\sigma$ is the Lipschitz constant of the function to be integrated and ${\rm diam}(P_j)$ is the maximum angular distance between any two points in pixel $j$.

In a continuously differentiable function the Lipschitz constant is the amplitude of the maximum derivative at any point. Several of our patterns are \textit{not} continuously differentiable due to the sharp horizon cut. This will be a very small effect for patterns that taper strongly towards the horizon, but it must be kept in mind for some patterns such as the monopole or very wide Gaussians.

Note that this approach gives us a pessimistic error bound -- it is essentially the bound one obtains by finding the pixel with the highest error and applying this error to all pixels and all having the same sign. Due to the oscillatory fringe, our errors should largely cancel from pixel-to-pixel, keeping errors well within these bounds.

The maximum diameter of the pixels can be obtained with any \code{healpix} implementation, for example with the function \verb|max_pixrad| in \code{healpy}, and is given to within 1\% for $N_{\rm side}>16$ by  $\underset{1\leq j \leq N_{\rm pix}}{\rm max} \left\{ {\rm diam}(P_j)\right\} = 7.4/\sqrt{N_{\rm pix}}$.

Our error bound is then
\begin{equation}
|I_{\rm true} - I_{\rm num}| \lesssim 2 \pi \frac{7.4\: \underset{l,u}{\rm max} |\nabla f|  }{\sqrt{N_{\rm pix}}}.
\label{eq:err_approx0}
\end{equation}

For application to the RIME, recall that the tangential gradient on the sphere (with the coordinates as defined in \S\ref{sec:definitions}) is
\begin{equation}
    \nabla = \frac{\hat{\theta}}{\sin\phi} \frac{\partial }{\partial \theta}  +  \hat{\phi} \frac{\partial }{\partial \phi}.
\end{equation}
Note that this is undefined at zenith (i.e. $\phi=0$). While in principle we could obtain the gradient at zenith via a different choice of coordinates, we find that the gradient there is almost never the maximum, due to sky symmetry, and we thus ignore it in this work.

The real part of the integrand is\footnote{The imaginary part is similarly defined, with $\sin$ instead of $\cos$.}
\begin{multline}
    f(\theta,\phi) = \frac{I(\theta,\phi)}{\cos\phi} \cos[2\pi (u \sin\theta\sin\phi 
    + v \cos\theta\sin\phi \\ + w(1 - \cos\phi)].
    \label{eq:rime_integrand}
\end{multline}
In App.~\ref{app:maxgrad} we find that, while a general maximum gradient cannot be determined exactly, for circularly symmetric patterns and with $w=0$, we have
\begin{multline}
    |\nabla f| = \left|\frac{\partial I}{\partial l} \cos (2 \pi  u l) \right. \\
    +\left. I(l) \left[\frac{l}{1-l^2} \cos (2 \pi  u l )-2 \pi  u \sin (2 \pi  u l)\right]\right|.
    \label{eq:symm_err_bound}
\end{multline}

 For reasonably smooth $I(l)$ that decreases faster than $1-l^2$, and when $u \gtrsim 1/2\pi$, only the last term of \cref{eq:symm_err_bound} contributes. Inserting this into \cref{eq:err_approx0} leaves
\begin{equation}
    |I_{\rm true} - I_{\rm num}| \lesssim \frac{29.6\pi^2}{\sqrt{N_{\rm pix}}} I(r_{\rm max}) u, \ \ \ r_{\rm max} \approx 1/4u.
\end{equation}
Since $u$ is reasonably large, and $I(l)$ is wide, $I(r_{\rm max}) \approx 1$.
For peaky patterns with a fast rate of change, the maximum may be dominated by the first term of \cref{eq:symm_err_bound}, in which case it occurs at close to the maximum of the derivative (again, if $u$ is reasonably large). Thus we can approximate the error bound as
\begin{equation}
    |I_{\rm true} - I_{\rm num}| \lesssim \frac{14.8\pi}{\sqrt{N_{\rm pix}}} \max_{u} \left\lbrace \max_{r} \{I'(r)\}, 2\pi u \right\rbrace
    \label{eq:errorbound}
	 \end{equation}

If $u \lesssim 1/2\pi$, the maximum is much more difficult to obtain. Nevertheless, we can say that it will not depend significantly on $u$. Thus, the overall picture as a function of $u$ is that at the smallest $u$ the error is essentially flat and governed by the sharpness of the sky model. At high $u$, the error bound increases linearly with $u$, independently of the sharpness of the sky model.

\section{Numerical Integration vs. Theory}
\label{sec:comparison}

For each pattern presented in \S\ref{sec:analytic_solns}, we generate \code{healpix} maps which we provide to \code{healvis} to produce simulated visibilities.
We make three \model{projgauss} and \model{gauss} maps using  width parameters $a=$0.05, 0.5, and 0.25, which we refer to as the ``small'', ``medium'', and ``large'' \model{projgauss}/\model{gauss} patterns, respectively. We also generate the \model{xysincs} pattern with the parameter $a = 64$ and the angle $\xi = \pi / 4$. For each pattern, we construct maps with $N_{\rm side}$ parameters of 256, 512, and 1024, to explore the behavior of errors with resolution.

We choose for our simulated array a configuration which shares the densely packed drift-scanning features common to many \HI cosmology low frequency arrays. Our fictional array has 128 antennas distributed according to a radially-Gaussian profile with a FWHM of 228~m. The spectrum spans 10 frequency channels in steps of 80~kHz starting at 100~MHz. This provides a good UV sampling that is consistent across simulations, and especially on scales near $|u| \ll 1$ that are most sensitive to wide-field structures. For this layout, $w=0$ on all baselines; this restriction is relaxed later.

The simulated array is located on the equator with zenith pointing at RA = Dec = 0; on-zenith patterns are evaluated centered on this point. The \textsc{healpix} map longitude and latitude are interpreted as RA and Dec, respectively. For each \model{gauss} width, additional simulations were run with the center of the pattern offset from zenith by either $0.5^\circ$ (``small-offzen''), $5.0^\circ$ (``offzen''), or $50.0^\circ$ (``large-offzen'') in longitude.

\subsection{Results}

\begin{figure*}[ht!]
\includegraphics[width=\linewidth]{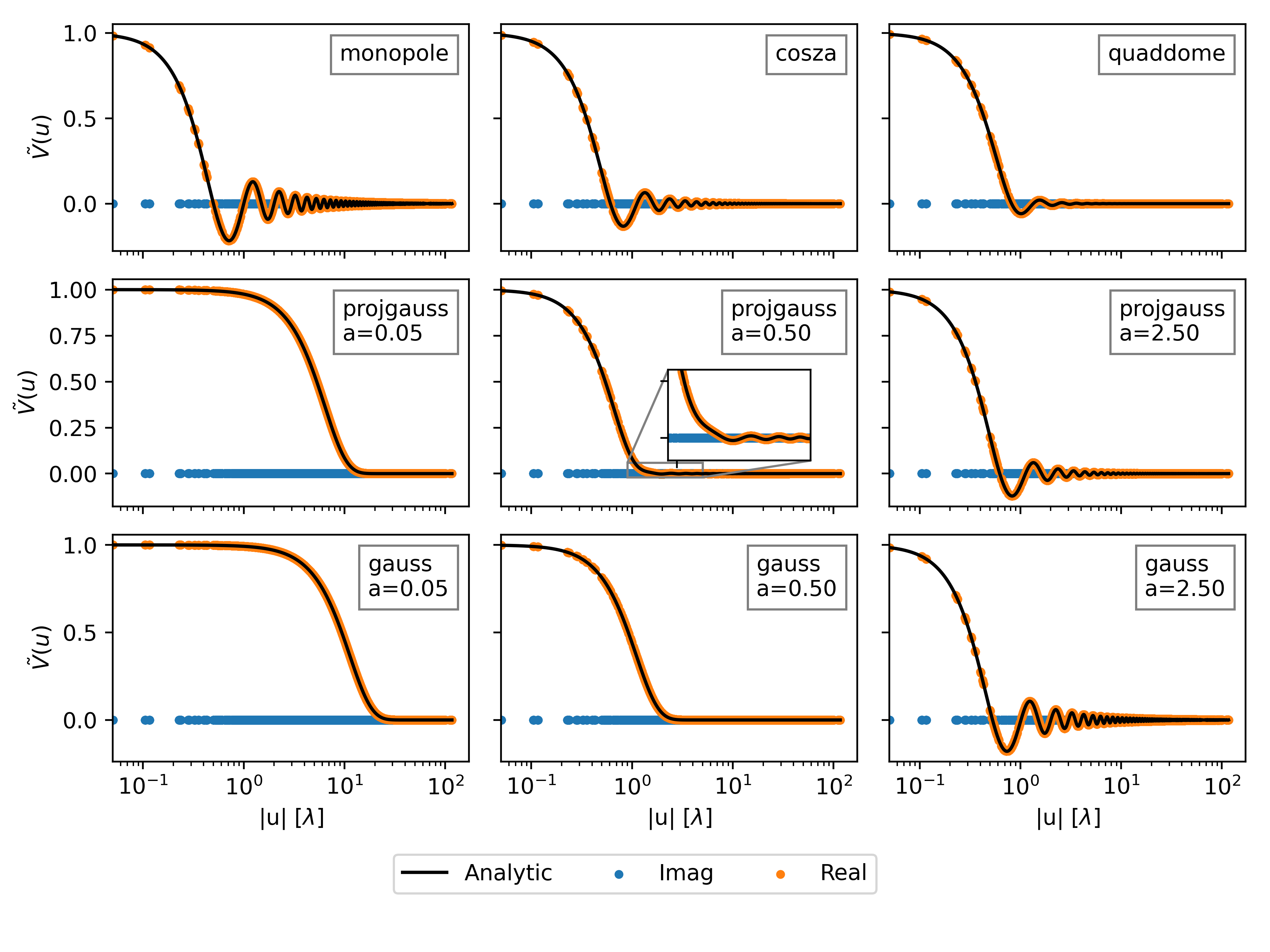}
\caption{Analytic 1D solutions (black) vs. real (orange) and imaginary (blue) components of axially-symmetric pattern simulations. Simulations used maps at $N_{\rm side}$ = 1024.
}
\label{fig:allsoln_compare}
\end{figure*}

\Cref{fig:allsoln_compare} compares visibilities simulated from $N_{\rm side}$ = 1024 maps to the analytic solutions for the axially-symmetric patterns. Visibilities are normalized as $\tilde{V} = V(|u|)/V(0)$, to the analytic value at $|u| = 0$, to account for the different overall amplitudes of the patterns. Since these are symmetric patterns, the imaginary component (shown in blue) is close to zero.

\begin{figure}[ht!]
\centering
\includegraphics[width=\linewidth]{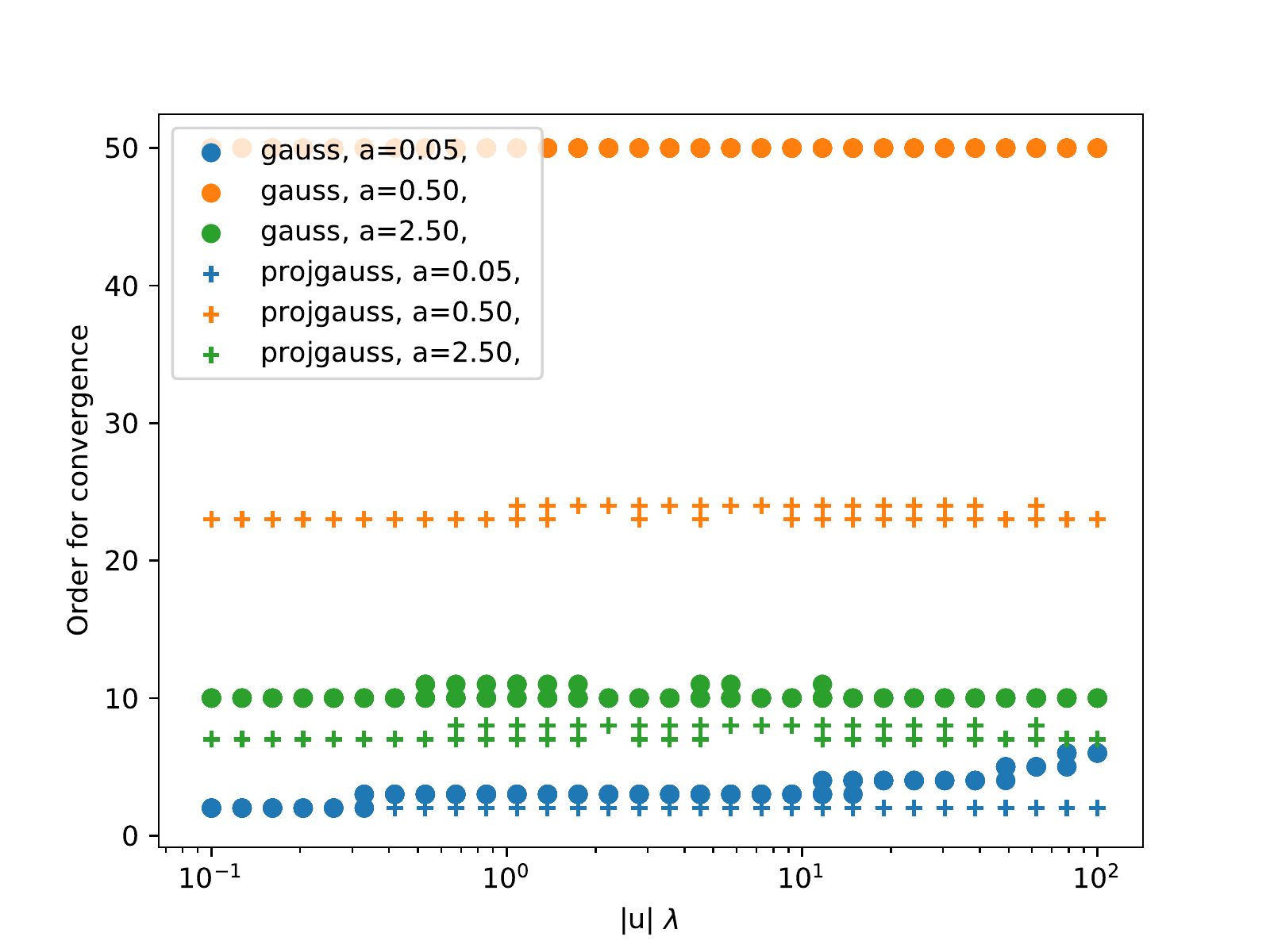}
\caption{The visibility solutions to the Gaussian (on-zenith) or projected Gaussian patterns use series expansions which must be tested for convergence. Here we show how many terms are needed until subsequent terms vary within machine precision. Throughout this section, series solutions are evaluated to 60 terms, which is sufficient to ensure convergence. Though not shown in this figure, this order is also sufficient for convergence for the off-zenith Gaussian cases as well.}
\label{fig:convergence_orders}
\end{figure}

As we found in \S\ref{sec:analytic_solns} the convergence of series solutions depends somewhat on the array configuration. Given a specific size of test pattern (e.g. selection of Gaussian $a$) and baseline distribution we can find the expansion order needed to converge.
\Cref{fig:convergence_orders} shows the minimum number of terms needed before subsequent terms different to order of machine precision. In the worst case of a large Gaussian at zenith, 50 terms were needed.  This case ($a=0.5$) is on the cusp between the small-$a$ and large-$a$ series; it is not surprising that it is the slowest to converge.  Throughout our analysis here \model{gauss} and  \model{projgauss} analytic solutions are evaluated to order 60.

With both Gaussian patterns, there is a clear transition in behavior when shifting between the small-a and large-a expansions. The inset in the center plot of \cref{fig:allsoln_compare} zooms in on the region just beyond the main Gaussian of the \model{projgauss} solution, showing jinc-like oscillations. A similar transition can be observed for medium-scale \model{gauss} results, but it is not apparent for $a=0.5$. At $a = 2.50$, the  \model{projgauss} pattern strongly resembles the \model{cosza} results, which is consistent with our expectation from the limiting behavior of \cref{eq:projgauss:largesig}. Since the guassian only goes to zero at infinity, the pattern will be non-zero at the horizon. This sharp transition manifests in ringing of the visiblities which grows with the width.

The simulated visibilities do appear to match the analytic results very well by eye. In the next subsections, we will examine this difference more closely.

\subsection{Error Analysis}

\begin{figure*}
\includegraphics[width=\linewidth]{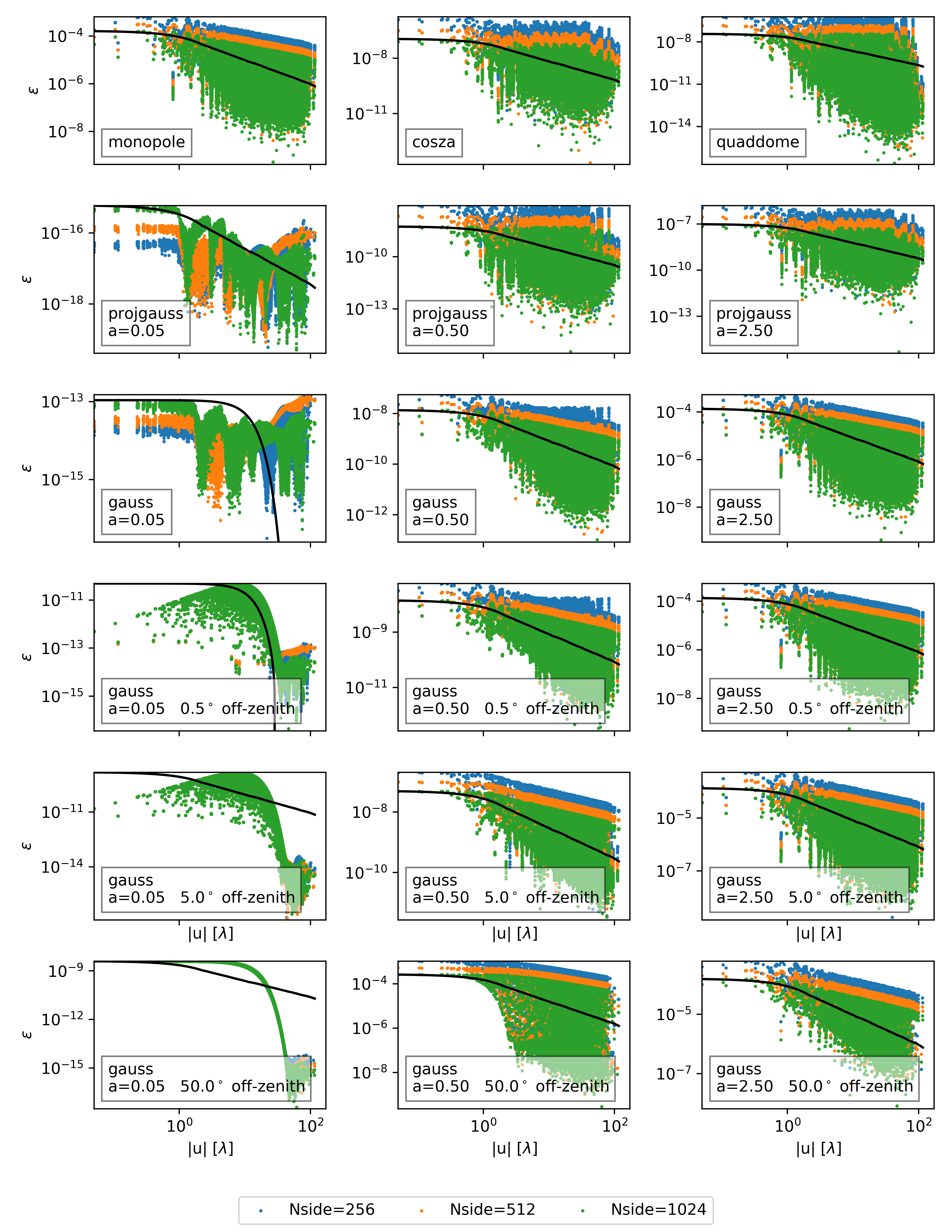}
\caption{Errors vs. baseline length and resolution, for the highest resolution simulations, of the patterns in \cref{fig:allsoln_compare} as well as the off-zenith \model{gauss} patterns. The overall amplitude of the errors drops off with increased resolution, except for the a=0.05 \model{projgauss} and \model{gauss} cases. For those, the errors are so low ($\lesssim 10^{-11}$) that it is likely floating point errors dominating over remaining simulation errors. The black lines are the visibility envelopes discussed in \S\ref{sec:1derrors_bllength}.}
\label{fig:1derrors}
\end{figure*}

In this section, we will aim to see if the remaining errors are sensible given our understanding of how PSA errors should behave -- Increasing roughly linearly for long baselines, and decreasing at least as fast as $1/\sqrt{N_{\rm pix }}$ for different resolutions. Exploring the deviations between the analytic solutions and simulation results can reveal subtle systematic issues in the integration scheme. Although the PSA is not necessarily the optimal approach to carrying out the RIME integral, it will serve as a demonstration of the utility of these test patterns.

\begin{figure}
\centering
\includegraphics[width=\linewidth]{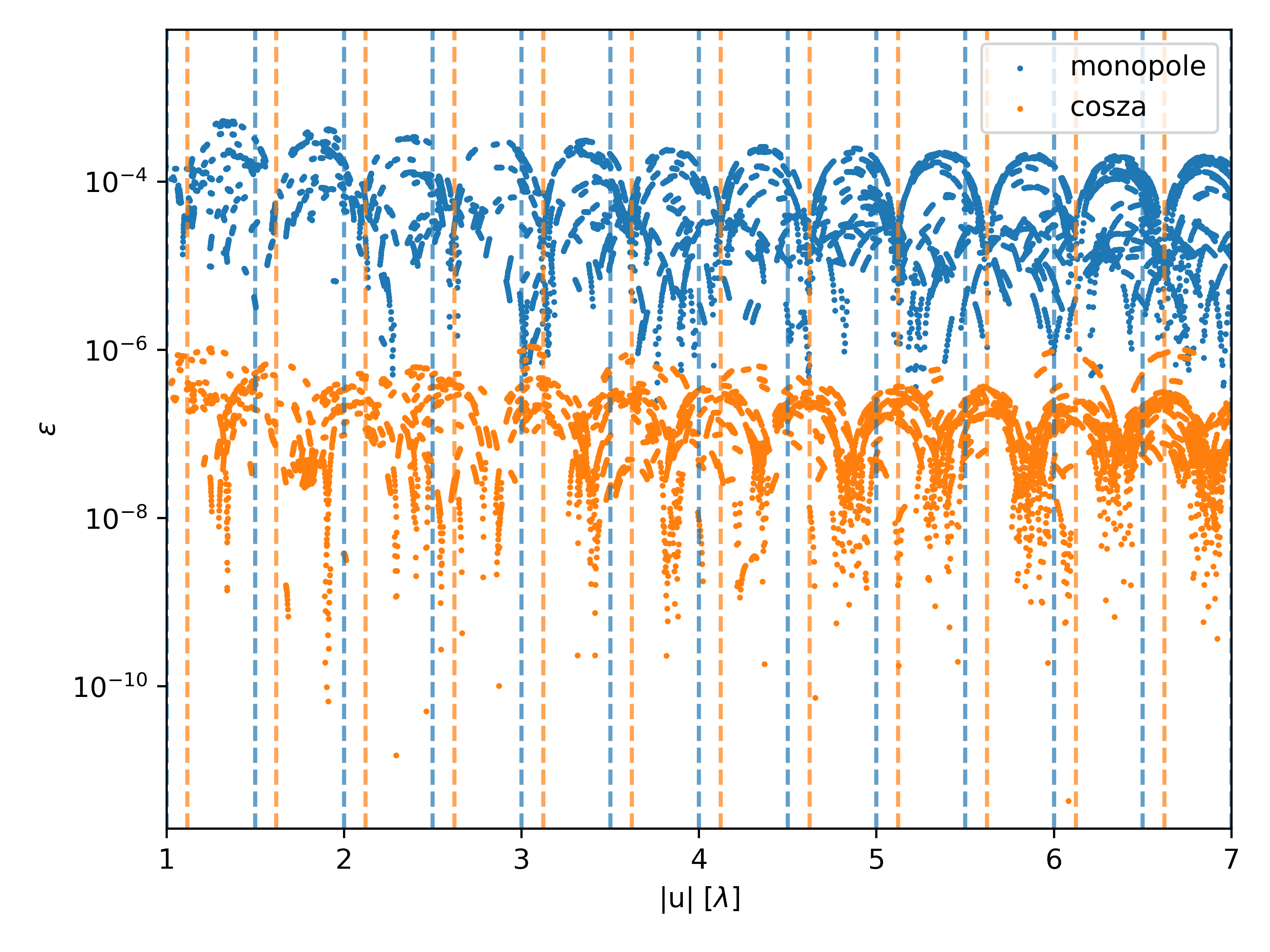}
\caption{A closer view of integration error for the \model{monopole} and \model{cosza} patterns shows that they are periodic. Vertical lines mark the zeros of the sinc and Bessel functions. }
\label{fig:monoresid_periodic}
\end{figure}

\begin{figure*}
\includegraphics[width=\linewidth]{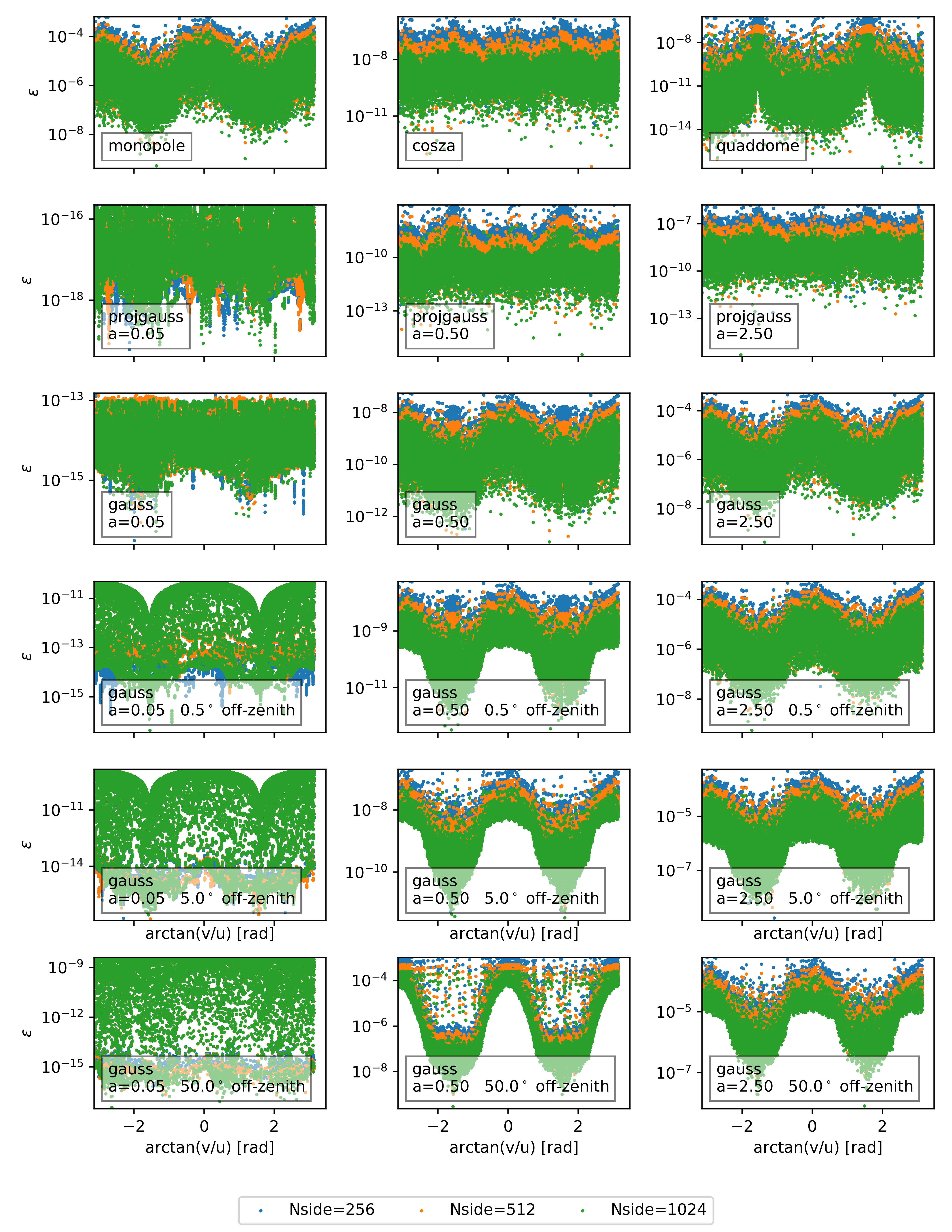}
\caption{Integration error vs. baseline angle in radians North of East, for the highest resolution simulations, of the patterns in \cref{fig:allsoln_compare} as well as the off-zenith \model{gauss} patterns. Compare with \cref{fig:1derrors}. For some models there is a noticeable angle dependence, which is apparently independent of baseline length and map resolution.}
\label{fig:1derrors_ang}
\end{figure*}

We define the error between numerical visibility simulation, $\tilde{V}_s$, and analytic solutions, $\tilde{V}_a$, as
\begin{equation}
\varepsilon(\vecb{u}) = |\tilde{V}_s(\vecb{u}) - \tilde{V}_a(\vecb{u})| .
\label{eqn:residual}
\end{equation}
where the visibilities are normalized to the analytically-calculated value at $|u| = 0$. \Cref{fig:1derrors} shows the errors (\cref{eqn:residual}) vs. baseline length for each symmetric model, for the three highest $N_{\rm side}$ values (i.e. finest resolution). Note that the $y$-axis range varies between plots by up to 7 orders of magnitude. \Cref{fig:1derrors_ang} shows the relative errors vs. baseline angle for the same resolutions.


\subsubsection{Baseline Length}
\label{sec:1derrors_bllength}
With oscillatory fringes, sharp horizon transition, and projection effects the integrand has several opportunities for strong dependence on baseline length. While the error bounds predicted by \cref{eq:errorbound} increase with baseline length, this calculation neglects the fact that the integral is highly oscillatory for large $u$. Examining $\varepsilon$ in \cref{fig:1derrors} we see that error drops off with increasing baseline length, or remains mostly constant depending on the model. 



It is difficult to visualise fractional error for a strongly oscillating function like the \model{monopole}, \model{cosza}, \model{quaddome}, and medium/large \model{gauss} and \model{projgauss} solutions, since the relative error is large when the solution crosses zero. One useful comparison is between the fractional error and the visibility magnitude; this makes clear when fractional errors are dominated by a shrinking denominator. The visibilities oscillate strongly with baseline length. In Fig.~\ref{fig:1derrors} we show the visibility ``envelope'' in black calculated by finding local maxima of the 1024 $N_{\rm side}$ visibility, then fitting by a piecewise linear function. In all cases, except for those with extremely small errors (e.g., narrow gaussians with $a=0.05$), the errors drop off more slowly than the visibility function; in other words, the fractional error slowly increases with baseline length.

Some errors vary periodically with baseline length, independently of map resolutions. Since this is difficult to see in \cref{fig:1derrors}, \cref{fig:monoresid_periodic} shows the \model{monopole} and \model{cosza} pattern errors, with a linear scale on the x-axis, over a limited range of baseline lengths. This error may be due to some small phase offset in the fringe term in the simulator, effectively shifting the set of baseline lengths off by some fixed small amount. Essentially, the sinc function of simulated data lags behind the analytic solution, creating a periodic offset (likewise for the \model{cosza} pattern).
This demonstrates the utility of these analytic solutions for identifying small systematic simulator errors such as these.

\subsubsection{Baseline Angle}
One important area to validate widefield simulators is azimuthal dependence, which we test here by varying baseline orientation while holding $|u|$ constant.  Azimuthally symmetric patterns should have the same solution on any baseline orientation, while patterns with position offsets or other rotational asymmetry will have similar asymmetry in Fourier space as they do on the sky.

\Cref{fig:1derrors_ang} shows the error vs. baseline angle, defined as the angle North of East (under the convention that the $u$-axis points East and the $v$-axis North), at the three highest resolutions. These are the same data points as in \cref{fig:1derrors}. There is a noticeable angular dependence in the errors for some, especially for the \model{monopole}, \model{gauss}, medium \model{projgauss}, and \model{quaddome} patterns.


This angle dependence almost certainly arises due to the discretization of \textsc{healpix} breaking the detailed angular symmetry of the pattern. The error is generally worse for East-West aligned baselines than for North-South baselines. Since our test pointing is on the \textsc{healpix} map equator, an East-West aligned baseline will be pointing along an isolatitude ring, which is uniformly divided in angular coordinates, while the North-South direction is more irregularly covered. Further simulations at different latitudes might shed some light on this effect.

\subsubsection{Map Resolution}
\begin{figure}
\centering
\includegraphics[width=\linewidth]{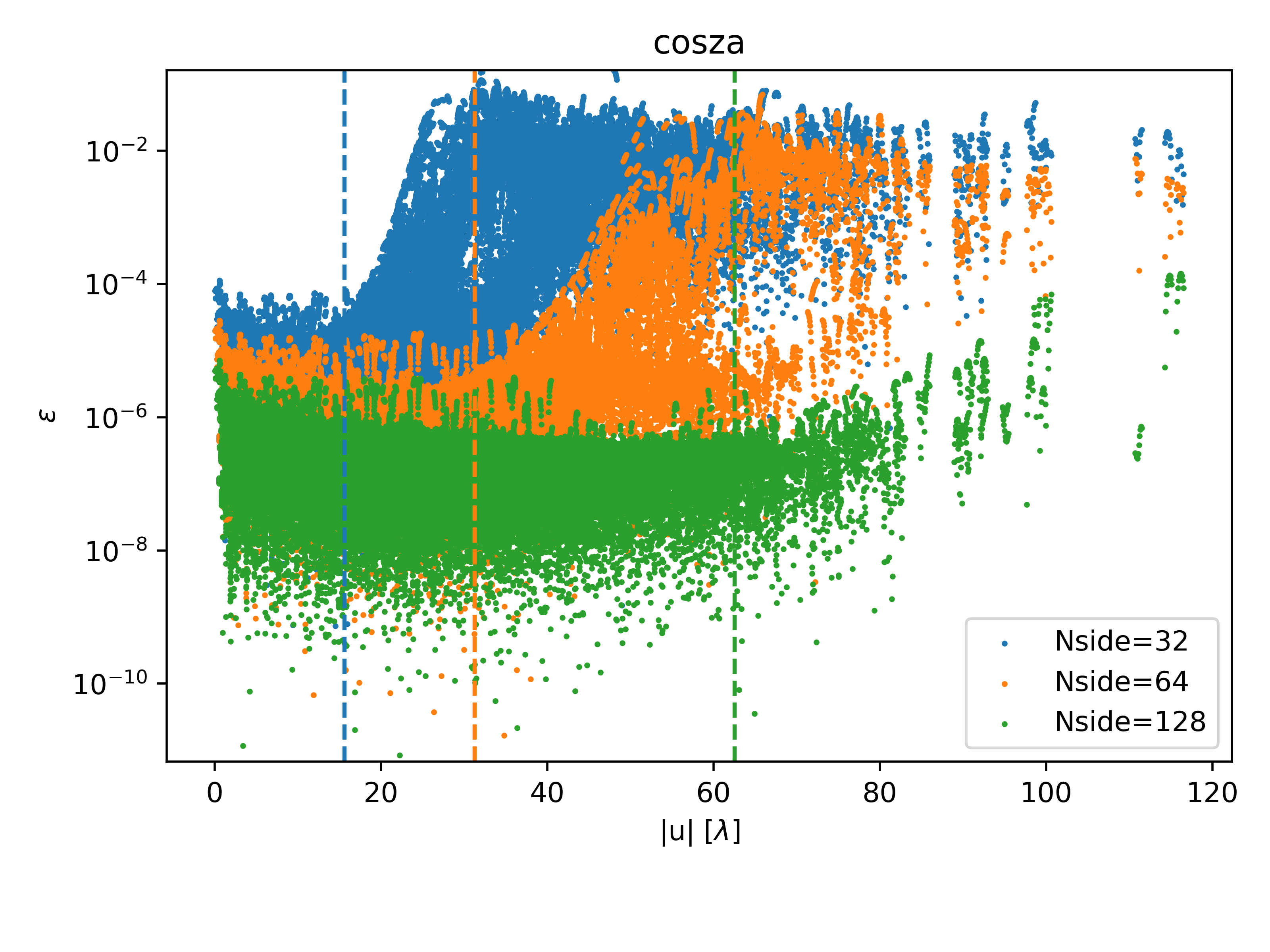}
\caption{An illustration of the impact of sampling density on the numerical integrator using the \model{cosza} pattern as an example. Error declines with increasing baseline until the nyquist point $u = 1/(2\sqrt{\Omega})$ (for pixel area $\Omega$, dashed lines) at which point the trend reverses. The effect becomes less prominent at denser samples. }
\label{fig:lowres_cosza}
\end{figure}

As the resolution of the map is increased in \cref{fig:1derrors} the overall trend is for error to decline, but there are some deviations from this. The errors of the \model{monopole} pattern are $\propto 1/N_{\rm side} \sim 1/\sqrt{N_{\rm pix}}$, which corresponds with the pixel length. The \model{cosza} and \model{quaddome} patterns, however, have errors that decrease as $1/N_{\rm side}^2$, corresponding with the pixel \emph{area}. We identified these trends by manually shifting the different color points by factors of $1/N_{\rm side}^2$.

A similar scaling is observed for the \model{gauss} and \model{projgauss} patterns. 
The small \model{projgauss} pattern has the lowest errors of all patterns, which may explain why it breaks the trend of decreasing error with increasing resolution. It is possible that the pixellization and PSA errors are so low that the error is dominated by machine noise, which scales roughly as the number of pixels $N_{\rm pix}$. Higher resolution patterns, which have more pixels, will therefore have slightly higher total error. The other \model{projgauss}  patterns have errors that scale as $1/N_{\rm side}^2$, faster than the theoretical bound $\propto 1/\sqrt{N_{\rm pix}}$. According to our estimate in \S\ref{sec:errorbound} we expect the upper performance bound to scale with resolution as $1/\sqrt{N_{\rm pix}}$ . Better scaling than this is not unexpected, if sky models happen to favourably match the regularity of the \textsc{healpix} sampling. 

Similarly, the small \model{gauss} simulations show extremely small error that doesn't vary much with resolution. They too, at $10^{-13}$, are likely hitting the machine noise floor. The medium and large \model{gauss} patterns show errors that scale approximately as $1/\sqrt{N_{\rm pix}}$. The large case strongly resembles the monopole pattern's errors. The errors on the off-zenith simulations are encouragingly similar to the on-zenith case, indicating that our off-zenith solution works as well as the on-zenith.

At very low resolutions ($N_{\rm sides}$ 32, 64, and 128), some simulated baselines are long enough that they over-resolve the pixels. \Cref{fig:lowres_cosza} shows the absolute error vs. baseline length using the \model{cosza} pattern as an example. Error declines with increasing baseline until $u = 1/(2\sqrt{\Omega})$ (for pixel area $\Omega$, dashed lines) at which point the trend reverses. This is approximately the Nyquist sampling rate, where the pixel is about half the fringe length.


\subsection{Check on rotational asymmetry with \model{xysincs}}

The \model{xysincs} test pattern explicitly breaks continuous rotational symmetry, allowing us to check for effects arising from orientation of large structure. It is also notable for its sharp edged pattern in $uv$ space which provides another useful visualization and check on baseline dependence.

\begin{figure*}
\includegraphics[width=\linewidth]{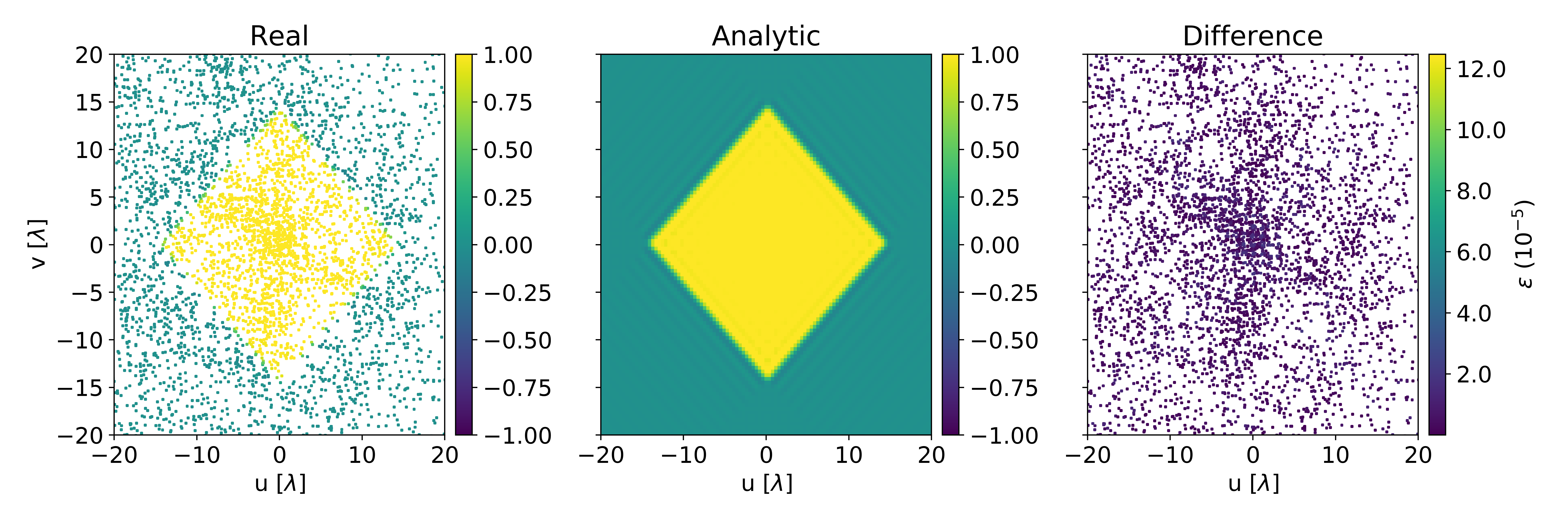}
\caption{\model{xysincs} pattern comparing real part of simulated data (left) from an $N_{\rm side}$  1024 map, analytic solution (centre), and the difference between the two at each simulated UV coordinate (right).}
\label{fig:xysincs_2dcompare}
\end{figure*}

\begin{figure*}
\centering
\includegraphics[width=0.45 \linewidth]{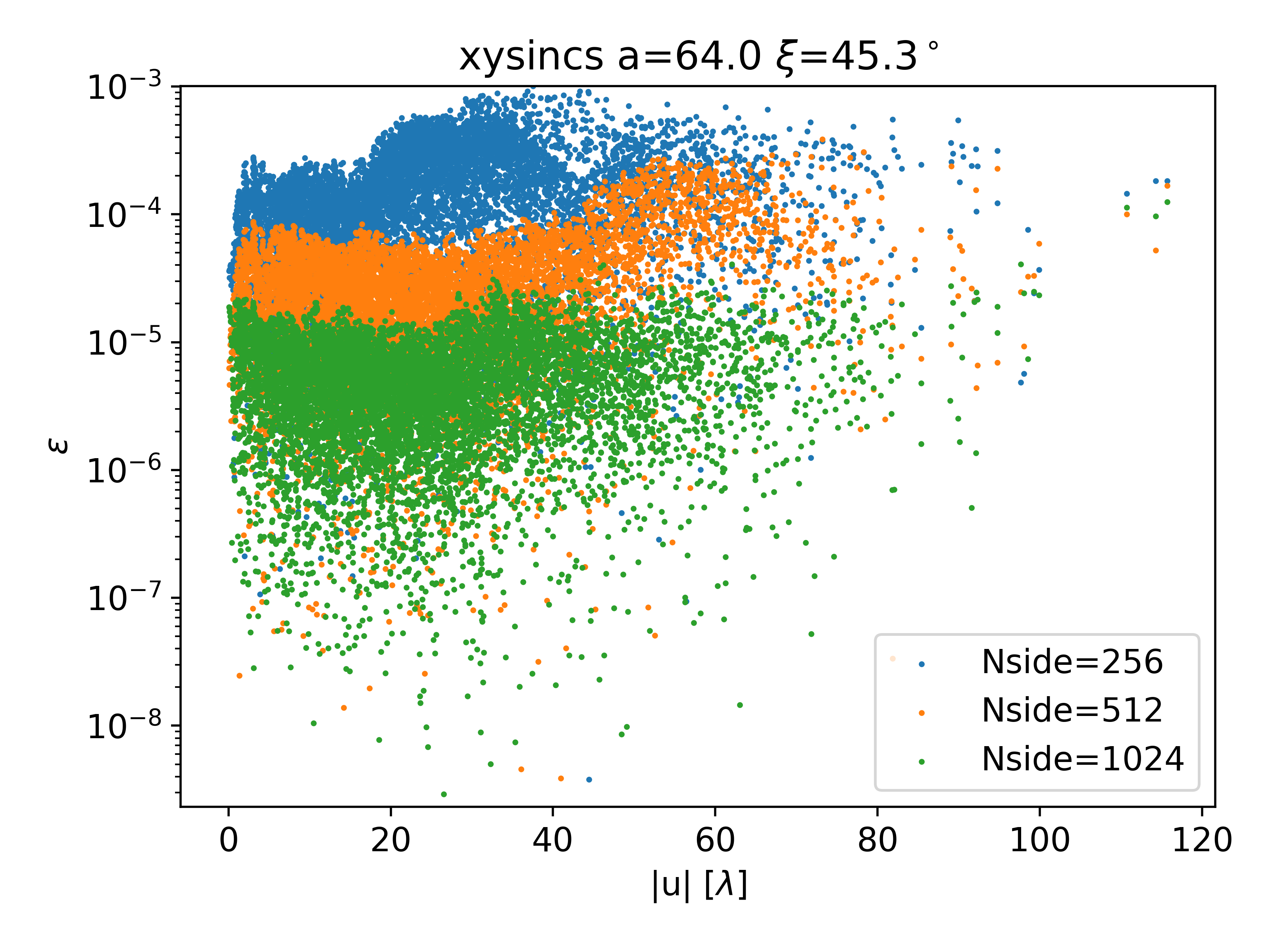}
\includegraphics[width=0.45 \linewidth]{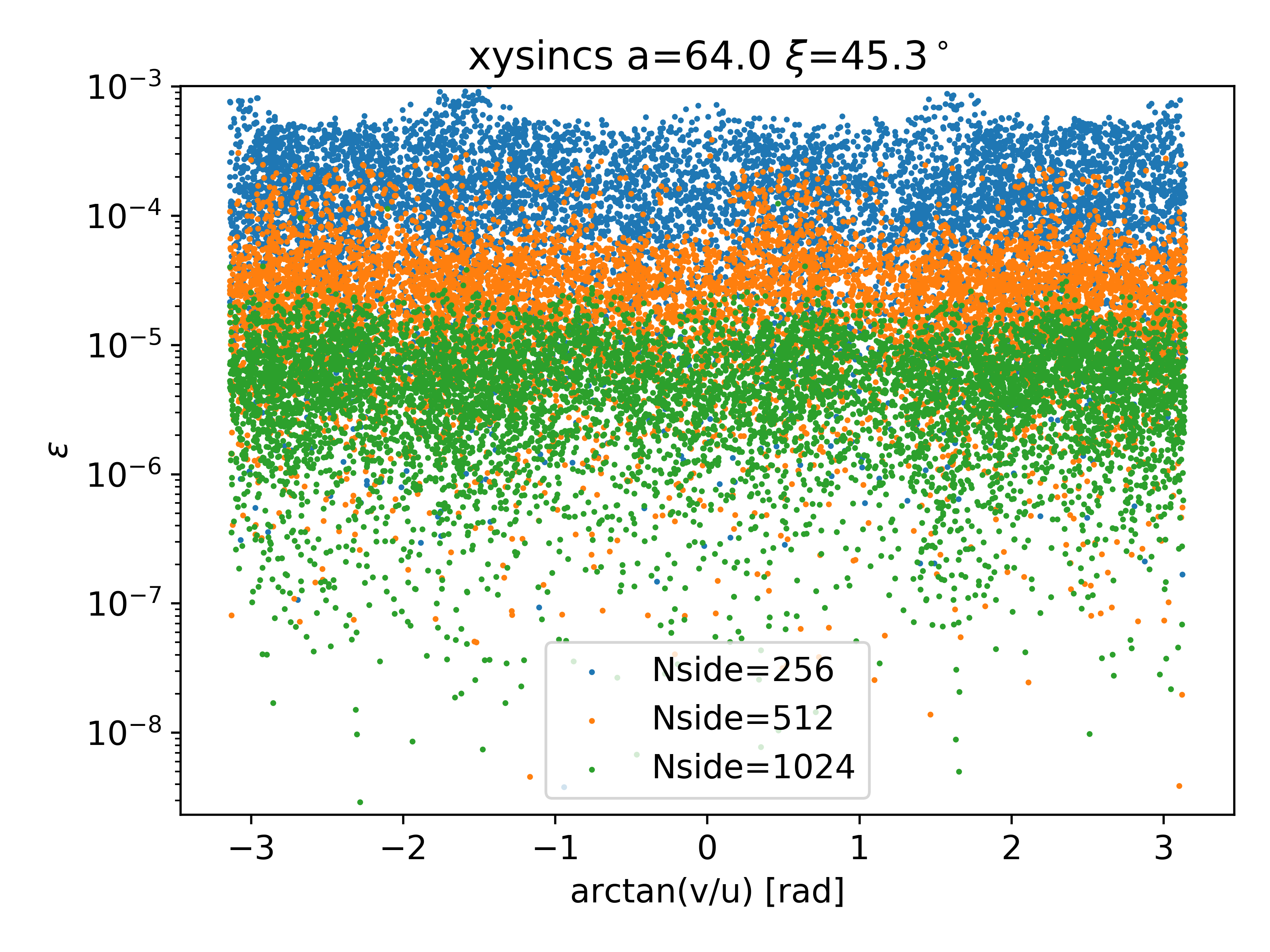}
\caption{Error of the \model{xysincs} pattern simulations at the three highest resolutions, vs. baseline length (left) and vs. baseline angle (right). Error drops off with increasing resolution, and doesn't show any noticeable dependence on angle despite the rotational asymmetry of the pattern.}
\label{fig:xysincs_errs}
\end{figure*}

\Cref{fig:xysincs_2dcompare} shows a side-by-side comparison of the normalized simulated visibilities, the analytic solution, and their difference for the \model{xysincs} pattern with $a=64$ and $\xi = 45^\circ$. The errors are of order $10^{-4}$, but don't show any noticeable dependence on $(u,v)$ position. This is shown more explicitly in \cref{fig:xysincs_errs}, which plots the absolute difference between simulated and analytic visibilities for \model{xysincs} patterns with the three highest resolutions. There is very little angular dependence, and the overall error scales roughly as $1/N_\text{pix}$. The upper bound on the error is comparable to that of the \model{monopole} test. Since the \model{xysincs} pattern has a sharp spatial cutoff (see~\cref{eq:separable_I}), we suspect the large error is due to the failure of the PSA to handle strong emission near the horizon. For the \model{monopole}, this is the actual circular horizon at $|\vecb{l}| = 1$, and for the \model{xysincs} this is the artificial square horizon introduced in \cref{eq:separable_I}.

\subsection{Non-Coplanar Baselines}
Another assumption we have made up to this point is that $w=0$; the approximation that all antennas lie in a plane.  However, not only are real arrays not perfectly flat, but the $w$ phase term multiplies the cosine of the elevation ($\sqrt{1-l^2-m^2}$) which is large at wide field angles. This assumption was necessary to arrive at a solution for all patterns except the monopole.

\begin{figure*}
\centering
\includegraphics[width=0.8\linewidth]{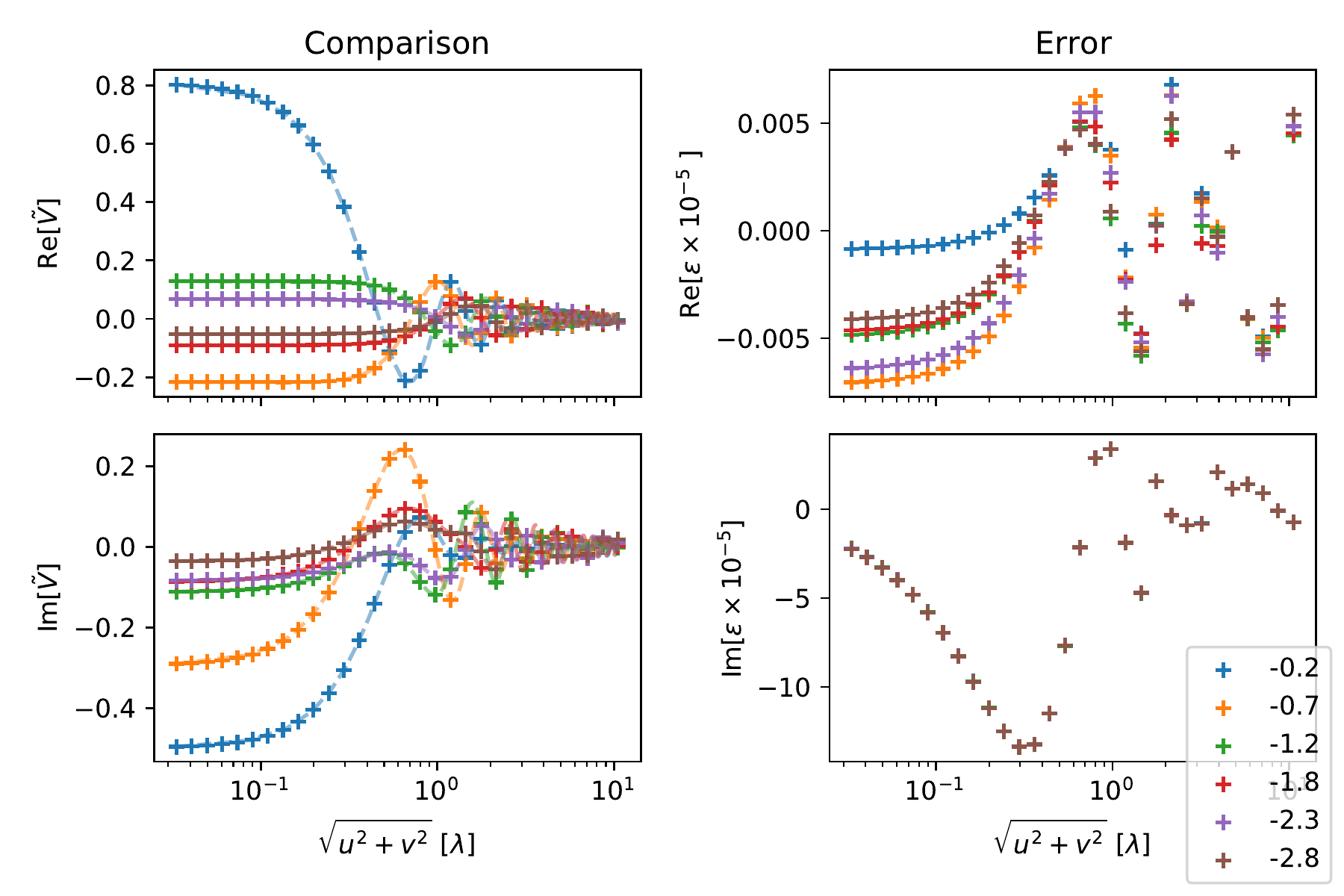}
\caption{Deviations from co-planarity increase integration error in monopole simulation, seen  here using $N_{\rm side}=1024$ with color indicating $w$ (in wavelengths). Analytic visibilities (dashed lines) and numerical visibilities (crosses) vs planar baseline length. Difference (right) in units of $10^{-5}$.}
\label{fig:noncoplanar_results}
\end{figure*}

Using the \model{monopole} pattern, we compute an additional set of simulations with non-zero $w$. Starting with a diagonal sample of baselines ($u=v$), with lengths spaced logarithmically from 0 to 22~m in 30 steps, we duplicate this set 20 times, each time adding a $w$ term 0 to 10~m. Altogether, this ``w-test'' layout comprised 600 baselines such that the solutions at fixed $(u,v)$ coordinates could be compared for different $w$ terms.\footnote{Realistic non-coplanarity does not extend to 10~m for typical \HI arrays. This w-test layout is designed to span a significant range of baseline lengths and $w$-amplitudes as a demonstration. Angular dependence is not important here as the sky brightness is azimuthally symmetric.}

\Cref{fig:noncoplanar_results} shows the results of simulations with the ``w-test'' layout observing a monopole sky at $N_{\rm side}=1024$, testing convergence under non-coplanar baselines. The left column compares the simulated to analytic visibilities, each curve for baselines with a different $w$ (in wavelengths). The series solution in \cref{eq:monopole} was evaluated to 80 terms.

The right column shows the difference between the simulated and analytic visibilities. These errors are all comparable to those seen for the coplanar monopole pattern at $N_{\rm side}=1024$. Interestingly, there is a consistent error in the imaginary component that does not seem to depend on $w$. This imaginary component dominates the real part error by about four orders of magnitude. Nonetheless, the errors are still very small, of order $10^{-5}$.

\section{Conclusions}
\label{sec:conclusions}

Visibility simulators have come to play an important role in the progress of \HI cosmology experiments -- testing analysis pipelines, calibrating data, and exploring the behavior of interferometers at the precision required to detect the 21cm background against foregrounds. It is therefore important that these simulators be tested at a level sufficiently more precise than the level needed in the experiment. When used for calibration, simulation products must be accurate to at least one part in 10,000. In general, simulation accuracy should be limited by the accuracy of the input data, and not by the algorithm used to evaluate the RIME.


In either case such testing benefits enormously from a fast and independently verifiable method, such as the analytic and series solutions to the visibility equation we have described here.  We have described a set of analytically-defined apparent brightness ``test patterns'' which reproduce several types of wide-field effects common to large, diffuse, sources spanning the curved sky to the horizon. For these test patterns we are able to evaluate the RIME integral to obtain closed-form or series visibility functions, which we call ``solutions''.  The solutions which require series expansion converge to machine precision with suitable number of terms. These patterns and their solutions may be used to test the precision and accuracy of simulated data in a way that requires no approximations to the RIME itself.  In searching for test patterns with analytic solutions we've aimed to cover a useful range of forms that mimic typical sky and beam configurations.

Among these solutions, we provide two forms of Gaussians which might be of general use for more traditional imaging and deconvolution. Under the common approximation of the RIME as a 2D Fourier transform (i.e., the flat-sky, narrow field of view approximation), Gaussians are often used to represent components of diffuse structures because they have well a well defined Fourier-transform (the Fourier-transform of a Gaussian being a Gaussian). We have shown that the solution for each Gaussian pattern does converge to a Gaussian, in visibility space, for narrow Gaussians on the sky. The relative error given by the leading order term in the series provides a scale of the error in taking the flat sky approximation (cf. App.~\ref{app:gaussian_smalla}).

We then illustrate their utility by evaluating the accuracy of a fiducial numerical visibility integrator, \code{healvis}. \code{Healvis} is optimized to simulate diffuse wide-field structure such as diffuse galactic foregrounds or the \HI background. We have generated \textsc{healpix} maps from the test patterns, and then numerically integrated them using \code{healvis}, finding it to be accurate to machine precision for some test patterns, but for others only to 1 part in 10,000. It seems to have particular difficulty with handling power near the horizon, and may include a persistent small phase offset whose cause is as yet undetermined.

Specifically, we observe the following trends in the errors:
\begin{itemize}
\item At low resolution, the errors spread to large values around where $|u| \sim 1/\sqrt{\Omega}$ for pixel area $\Omega$. This is the point where the fringe length is comparable to the pixel scale.
\item Error decreases with increasing $N_{\rm side}$ (smaller pixels), as expected. In all cases, the error decreases at least as quickly as $1/\sqrt{{\rm Npix}}$, which is the expectation for uniformly distributed random sample points. Since the HEALpix pixelization is regular, usually the errors drop off more quickly.
\item There is a slight angular dependence noticeable in some errors, which is likely due to the angular pattern in the  \textsc{healpix} pixel locations.
\item The periodicity of the errors with baseline length suggest that \code{healvis} has some small pointing error, and illustrates the importance of using exact solutions to uncover such small inaccuracies.
\item The largest errors are seen for patterns that are large at the horizon. This suggests that \code{healvis} has the most trouble handling pixels near the horizon, most likely due to the fact that some of the pixels extend beyond the sharp horizon cutoff.
\item There appears to be a lower bound on the error consistent with machine precision.
\end{itemize}

These questions can be probed further to increase the accuracy of integration using well-crafted spherical integration schemes, and more robust estimates of their error taking into account oscillations in the integrand. We leave these interesting questions for future work.

The simulated data generally show agreement with the solutions at one part in $10^5$ or better, a precision level sufficient for diagnostics of demanding \HI cosmology applications.  
In most cases the small discrepancies arise from discretization of the sky model into image pixels.
In practice, limited knowledge of the underlying sky-model covariances accumulated between observing and input into the simulator will contribute a much larger uncertainty. 
Such issues are only problematic if we aim to compare with data and irrelevant to the validation of simulators. A better understanding of sky model errors will be necessary to calibrate or subtract at the precision levels we report here.

We also recognize that implementation of these tests in a robust and repeatable way requires a significant investment of software infrastructure.  This is the motivation for the ongoing \code{pyuvsim} project which aims to provide a reliable simulator test product so instrument simulators can compare against a common reference standard. Both the point source approximation for diffuse simulations and the analytic tests in this paper have recently been incorporated into \code{pyuvsim}. The analytic tests are implemented as unit tests which guarantee long-term stability for codes testing against unintended changes. 

\section{Acknowledgements}
This research was suported by NASA Grant 80NSSC18K0389 for AEL, and used resources of the Center for Computation and Visualization (CCV) at Brown University. SGM and DCJ are supported by the HERA project under NSF grants AST1836019 and AST1636646. We are grateful to the developers and maintainers of \code{numpy} \citep{2020NumPy-Array}, \code{astropy} \citep{astropy:2013,astropy:2018}, \code{matplotlib} \citep{hunter_matplotlib}, \code{mpmath} \citep{mpmath}, \code{scipy} \citep{2020SciPy-NMeth}, \code{Wolfram Mathematica}, \code{jupyter} \citep{Kluyver2016jupyter}, and \code{pyuvsim} \citep{lanman_pyuvsim_2019}.

\bibliography{sample63}{}
\bibliographystyle{aasjournal}

\begin{appendices}

\section{Special Functions and their Identities}
\label{app:special_funcs}

\subsection{Complete and Incomplete Gamma}
\label{app:special:gamma}
The Gamma function can be thought of as an extension to the integer factorial function, and is defined as 
\begin{equation}
    \Gamma(z) = \int_0^\infty x^{z-1} e^{-x} dx,
\end{equation}
for all complex $z$ with positive real component. It can be extended to the entire complex plane (except for the negative real integers) by the identity
\begin{equation}
    \Gamma(z) = \Gamma(z+1)/z.
\end{equation}
For real integers, $z = k \in \mathbb{Z}$, the Gamma function reduces to the factorial:
\begin{equation}
    \Gamma(1+k) = k!.
\end{equation}

An important identity concerning the Gamma function that we use in this paper is Euler's reflection formula:
\begin{equation}
    \Gamma(1-z)\Gamma(z) = \frac{\pi}{\sin \pi z}.
\end{equation}
For integer $k$, setting $z=1/2-k$ yields the relation
\begin{equation}
    \Gamma(1/2 + k) \Gamma(1/2 - k) = \frac{\pi}{\sin\left(\pi (\frac{1}{2}-k)\right)} \equiv \pi (-1)^k.
    \label{eq:reflection_formula}
\end{equation}

The upper-incomplete Gamma function is defined as
\begin{equation}
    \Gamma(a, b) = \int_b^\infty t^{a-1}e^{-t} dt.
    \label{eq:inc_gamma}
\end{equation}

\subsection{Bessel Function}
\label{app:special:bessel}
The Bessel function of the first kind $J_\alpha(z)$ is a solution to Bessel's differential equation.
\begin{equation*}
z^2 \frac{d^2 f}{dz^2} + z \frac{d f}{dz} + (z^2 - \alpha^2) f = 0
\end{equation*}
has solution
\begin{equation*}
f(z) = J_\alpha(z)
\end{equation*}

$J_\alpha$ may be expressed as an integral:
\begin{equation*}
J_\alpha(z) = \frac{1}{2\pi}\int\limits_{-\pi}^\pi e^{i x (\sin t - n t)} dt
\end{equation*}

The modified Bessel functions $I_\alpha(z)$ are related to $J_\alpha(z)$ by
\begin{equation*}
I_\alpha(z) = i^{-\alpha} J_\alpha(i z) .
\end{equation*}

The modified Bessel function $I_0(z)$ has the Taylor expansion (about zero) of
\begin{equation}
    \label{eq:besselI_taylor}
    I_0(z) = \sum_{k=0}^\infty \frac{z^{2k}}{4^k\Gamma^2(1+k)}.
\end{equation}

\subsection{Hankel Transform}
The Hankel Transform of order $\nu$ of a function $f(r)$ can be defined as
\begin{equation}
    H_\nu(k) = \int_0^\infty r f(r) J_\nu(k r)dr,
\end{equation}
with $J_n$ the Bessel function of the previous section.

The Hankel Transform arises naturally when considering the Fourier Transform of a circularly symmetric function in $n$ dimensions:
\begin{align}
    F(k) &= \int d^n \vecb{r} e^{-2\pi i \vecb{k}\cdot \vecb{r}} f(|\vecb{r}|) \nonumber \\
    &= \frac{2\pi}{k^{n/2-1}}\int_0^\infty dr r^{n/2} f(r) J_{n/2-1}(2\pi k r) \\
    &\equiv H_{n/2-1}(2\pi k)[f(r)r^{n/2-1}]. \nonumber \\
\end{align}
In particular, in 2D, we have
\begin{align}
    \label{eq:fourier_to_hankel}
    F(k) &= 2\pi\int_0^\infty dr r f(r) J_0(2\pi k r) \\
    &\equiv 2\pi H_0(2\pi k). \nonumber \\
\end{align}

\subsection{Hypergeometric Function}
\label{app:special:hypergeometric}
The generalized hypergeometric function is defined as\footnote{\url{http://dlmf.nist.gov/15}}
\begin{equation}
    \pFq{p}{q}{a_1,\cdots,a_p}{b_1,\cdots,b_q}{z}
    = \sum_{n=0}^\infty \frac{(a_1)_n(a_2)_n\cdots (a_p)_n}{(b_1)_n(b_2)_n\cdots (b_q)_n} \frac{z^n}{n!},
    \label{eq:hypergeometric_def}
\end{equation}
with $(a)_n$ the Pochhammer symbol or ``rising factorial'':
\begin{equation}
(a)_n = a (a+1) (a+2) \cdots (a + n - 1) = \frac{\Gamma(a+n)}{\Gamma(a)}
\end{equation}

The expansion converges under certain conditions:
\begin{enumerate}
\item If $p \leq q$, the series converges for all $z$.
\item If $a_i$ is zero or negative for one or more of the $a$'s, then the $(a_i)_k$ is zero for all k's large than some $k$, so the series terminates and the function is a polynomial.
\item For $p \geq q+1$, the series may be redefined in terms of contour integration or analytic continuation. None of the examples considered here fall into this regime.
\end{enumerate}
In this text, three forms of hypergeometric function have come up -- ${}_0F_1$, ${}_1F_1$, and ${}_1F_2$. All three of these fall into the first category, and so are convergent.

The case of $p=0$ and $q=1$ are called the \emph{confluent hypergeometric limit functions}, and are related to Bessel functions of the first kind.
\begin{equation}
\pFq{0}{1}{}{\alpha+1}{-\frac{x^2}{4}}
= \Gamma(\alpha + 1)\left( \frac{x}{2}\right)^{-\alpha} J_\alpha(x)
\label{eq:confluent_hyper}
\end{equation}

With $p=1$ and $q=1$ we get the \emph{confluent hypergeometric functions of the first kind}, and is a solution to Kummer's differential equation:
\begin{equation*}
z \frac{d^2 f}{dz^2} + (b - z) \frac{df}{dz} - a f = 0
\end{equation*}
such that
\begin{equation*}
f(z) = {}_1F_1(a; b; z)
\end{equation*}

A particularly useful identity involving integrals over the zeroth-order Bessel function involves the hypergeometric function, viz.:
\begin{equation}
    \int_0^1 x^n J_0(x) dx = \frac{\pFq{1}{2}{(1+n)/2}{1, (3+n)/2}{ -\frac{1}{4}}}{1+n}, \ \ \ \ n \geq 0.
    \label{eq:integral_of_power_j0}
\end{equation}

\begin{figure}
    \centering
    \includegraphics{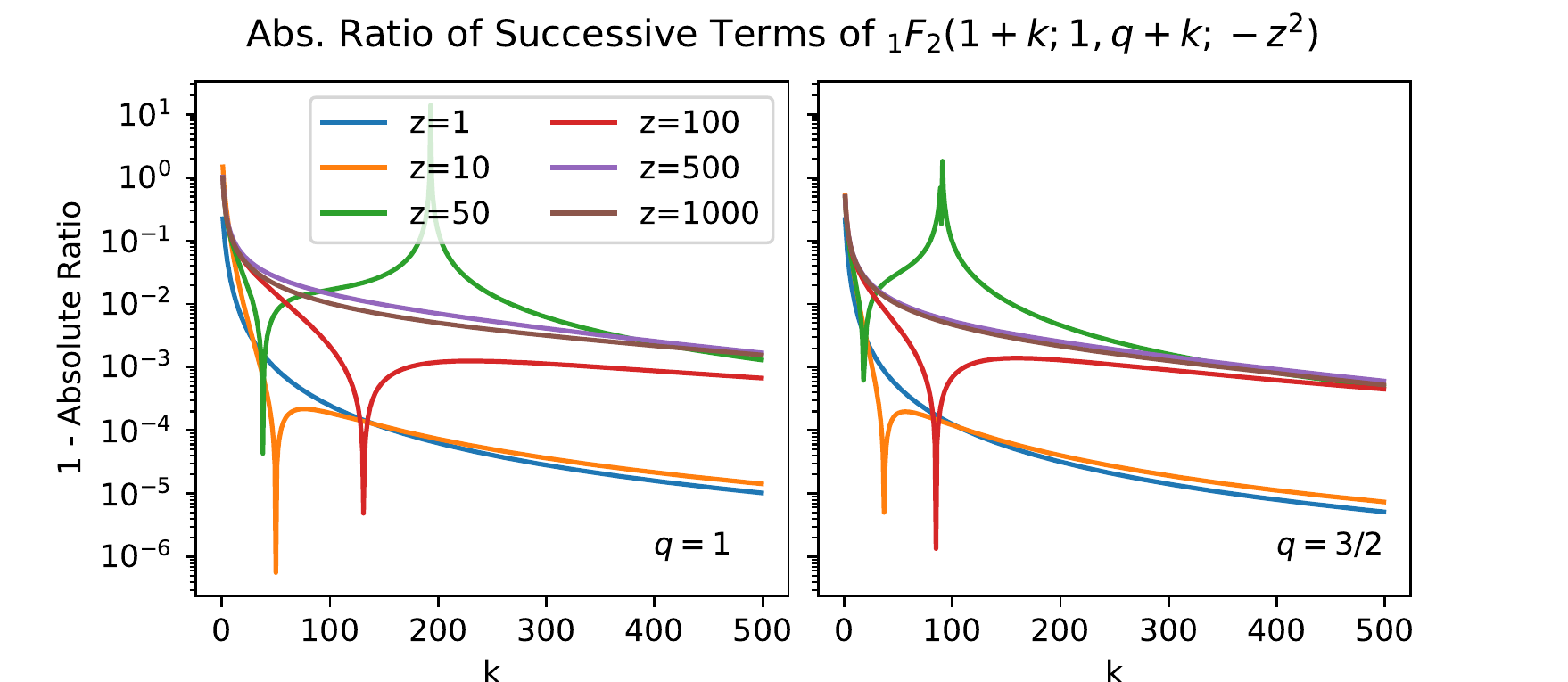}
    \caption{Absolute ratio of successive terms of $\Xi(k)$ for several realistic values of $z$ and the two values of $q$ encountered in solutions in this paper. The term ratios tend to a constant very close to unity, implying the function itself asymptotes to something very close to a constant.}
    \label{fig:hyp1f2_asymptotic}
\end{figure}

The function ${}_1F_2$ occurs several times in our solutions, and it thus is useful to understand some of its properties.
In particular, the function $\Xi(k|q,z) = \pFq{1}{2}{1+k}{1,q+k}{-z^2}$, with $q$ taking the values $q=1$ or $q=3/2$, occurs in the solutions to Gaussian and Projected Gaussian patterns. 
We have already noted that this function is itself absolutely convergent. However, it is important to understand the behaviour of this function for large $k$, as it typically appears inside a sum over $k$.
\cref{fig:hyp1f2_asymptotic} illustrates the asymptotic behaviour of the function as $k$ increases, for several realistic values of $z$. In the figure, we plot the value $1 - |\Xi(k+1)/\Xi(k)|$, i.e. the absolute ratio of successive terms. It is clear that these asymptote to a constant very close to unity.
Thus, summing this function over $k$ would not itself converge, but when multiplied by a strong decreasing function of $k$, it will converge absolutely.

\subsection{Sine Integral}
\label{app:special:sine}
The sine integral function is defined as
\begin{equation}
    \text{Si}(x) \equiv \int\limits_0^x \frac{\sin t}{t} dt
\end{equation}

\subsection{Laguerre Polynomials}
\label{app:special:laguerre}
The Laguerre polynomials have the closed form definition
\begin{equation}
    L_n(x) = \sum_{k=0}^n \binom{n}{k}\frac{(-1)^k}{k!} x^k, 
    \label{eq:laguerre_def}
\end{equation}
where we have used the binomial coefficient
\begin{equation}
    \binom{n}{k} \equiv \frac{n!}{k!(n-k)!}.
\end{equation}

Laguerre polynomials have the useful property 
\begin{equation}
    L_{-n}(x) = e^x L_{n-1}(-x).
    \label{eq:laguerre_neg_identity}
\end{equation}

\section{Compact Gaussian Approximation}
\label{app:gaussian_smalla}

To obtain a simpler estimate of the convergence of \cref{eq:gaussian_expansion}, 
we restrict ourselves to compact Gaussians, and use the asymptotic expansion \citep{paris2013}
\begin{equation}
    {}_1F_1(a,b,-|x|) = \frac{\Gamma(b)}{\Gamma(a)}x^{a-b}e^{-x}\left(1 + \mathcal{O}(1/x)\right) + \frac{\Gamma(b)x^{-a}}{\Gamma(b-a)} \sum_{n=0} \frac{\Gamma(a+n)\Gamma(1+a-b+n)}{x^n \Gamma(n+1)\Gamma(a)\Gamma(1+a-b)}.
    \label{eq:gauss_asymptotic_expansion}
\end{equation}
We have $a =1+k, b=3/2$, for which (for large values of $|x|$) the second term is extremely dominant at both small and large values of $k$, but in which the first term is dominant between these values. 
The range in which the first term is dominant begins at a minimum $k$ which depends positively on the value of $|x|$ according to
\begin{equation}
    k_{\rm mx} ( 2 \log(2k_{\rm mx}) + 4 + \log 4) \approx x - t \log 10,
\end{equation}
with $t$ an order-of-magnitude threshold for what constitutes ``dominant'' (eg. solve with $t=8$ to determine the maximum usable $k$ in which the second term dominates by 8 orders of magnitude).
As a rule-of-thumb, for $t=8$, with $a\lesssim0.1$, $k_{\rm mx}\gtrsim 50$, and we assume for the remainder of this section that this condition is true. We note that for $u a^2 \lesssim 2/3$ we have already established that the sum is convergent before $k=50$, and we will consider only this case in the remainder of the section.

Omitting the first term, and substituting our values in, we have
\begin{align}
    {}_1F_1(1+k; 3/2; -\frac{\pi}{a^2}) &\approx \frac{\Gamma(3/2)}{\Gamma(1/2-k)\Gamma(1+k)\Gamma(1/2+k)} \left(\frac{a^2}{\pi}\right)^{1 +k} \sum_{n=0}^{m} \left(\frac{a^2}{\pi}\right)^{n} \frac{\Gamma(1+k+n)\Gamma(1/2+k+n)}{\Gamma(n+1)} \nonumber \\ 
    &= (-1)^k \frac{1}{2\sqrt{\pi}\Gamma(1+k)} \left(\frac{a^2}{\pi}\right)^{1 +k} \sum_{n=0}^{m} \left(\frac{a^2}{\pi}\right)^{n}\frac{\Gamma(1+k+n)\Gamma(1/2+k+n)}{ \Gamma(n+1)},
\end{align}
where the second line uses Euler's reflection formula (cf. \cref{eq:reflection_formula}).

Inserting this expression back into \cref{eq:gaussian_expansion} and summing over $k$ gives
\begin{align}
    V_{\rm g}(u|a \lesssim 0.1) &\approx  \frac{a^2}{\sqrt{\pi}} e^{-\pi a^2 u^2} \sum_{n=0}^m \left(\frac{a^2}{\pi}\right)^n 
  \Gamma(1/2 + n) {}_2F_1\left(\frac{1}{2} + n, 1 + n; 1; -a^4 u^2\right).
  \label{eq:gaussian_expansion_approx}
\end{align}
The remaining hypergeometric function is convergent iff $a^2 u < 1$, which is a condition we have already assumed.

The dominant factor of the sum is the power-series in $a^2$, which for the small $a$ that we are considering drops off rapidly. Therefore relatively few terms are
required for convergence (eg. convergence is considerably faster than in \cref{eq:gaussian_expansion}).

The first few terms of the expansion can be written explicitly:
\begin{align}
    V_{\rm g}(u|a \lesssim 0.1) &\approx  \frac{a^2}{\sqrt{1 + a^4 u^2}} e^{-\pi a^2 u^2} \left(1 + \frac{a^2(-2 + a^4u^2)}{4\pi (1 + a^4u^2)^2} + \frac{3 a^4 (8 - 24 a^4 u^2 + 3 a^8 u^4)}{32 \pi^2 (1 + a^4 u^2)^4} + \mathcal{O}(a^6)\right)
\end{align}
Clearly, as $a\rightarrow0$, it yields the simple Gaussian solution.


Alternatively, we may estimate the first-order relative error induced by making the common assumption that the Fourier Transform of a Gaussian is itself a Gaussian:
\begin{equation}
    \epsilon \approx 1 - \frac{1}{\sqrt{1+u^2a^4}}. 
\end{equation}

Consider a set of baselines for which we would like to simulate visibilities given a sky model consisting of a single Gaussian blob that is barely resolved -- i.e. the blob has size $a_c = (2\pi)^{3/2}/u_{\rm max}$. 
The baseline which is simulated with the largest error will be the longest one, i.e. with a length $u_{\rm max}$, and thus the first-order error estimate has an upper bound of 
\begin{equation}
    \epsilon \approx 1 - \frac{1}{\sqrt{1+(2\pi)^6 u_{\rm max}^{-2}}}. 
\end{equation}
To even achieve 1\% error requires using baselines out to $u_{\rm max} \approx 1700\, \lambda$ (note that this still requires that $a_c < 0.1$, and thus  $u_{\rm max} \gtrsim 150$).
In this sense, to the level of accuracy generally required for analytic solutions, this assumption is never tenable for typical low-frequency observatories, and the more general forms of \cref{eq:gaussian_expansion} or \cref{eq:gaussian_expansion_approx} are required.

\section{Maximum Gradient of the RIME integrand}
\label{app:maxgrad}

We aim to find the maximum magnitude of the gradient of Eq. \ref{eq:rime_integrand}, i.e. $\underset{\theta,\phi}{\rm max} |\nabla f|^2$.
We have
\begin{align}
    |\nabla f|^2 = &\sec^2\phi \Bigg[\left( \frac{\partial I}{\partial \phi} \cos (2 \pi  g(\vecb{u},\theta,\phi)) -2 \pi  I(\theta ,\phi ) g_\phi(\vecb{u},\theta,\phi) \sin \left[ 2 \pi  g(\vecb{u},\theta,\phi)\right] + \tan\phi I(\theta ,\phi ) \cos \left[2 \pi  g(\vecb{u},\theta,\phi)\right]\right)^2 \nonumber \\
    &+\left. \left(\csc\phi \frac{\partial I}{\partial \theta} \cos (2 \pi  g(\vecb{u},\theta,\phi))+2 \pi  I(\theta ,\phi ) (u \sin\theta- v \cos\theta) \sin (2 \pi  g(\vecb{u},\theta,\phi))\right)^2\right],
    \label{eq:fullerrorbound}
\end{align}
with $g = \vecb{u}\cdot \hat{s} = u\cos\theta\sin\phi + v\sin\theta\sin\phi + w(1 - \cos\phi)$ and its derivative with $\phi$, $g_\phi = u\cos\theta \cos\phi + v\sin\theta\cos\phi + w\sin\phi$.

In general, it is unlikely that the maximum of this function may be found analytically for any particular brightness function $I$. However, it may be determined numerically via gradient descent algorithms. For the circularly symmetric patterns presented in \S\ref{sec:analytic_solns}, and with $w=0$, we can without loss of generality take the solution at $v=0$ and note that the direction of the greatest derivative then must be along the $l$ axis (since the fringe oscillates in this direction only, and the sky is symmetric). We thus obtain
\begin{align}
    |\nabla f| &= \left|\sec\phi \frac{\partial I}{\partial \phi} \cos (2 \pi  u \sin\phi)+I(\phi ) (\tan\phi \sec \phi \cos (2 \pi  u \sin\phi )-2 \pi  u \sin (2 \pi  u \sin\phi))\right| \nonumber \\
    &= \left|\frac{\partial I}{\partial l} \cos (2 \pi  u l)+I(l) \left[\frac{l}{1-l^2} \cos (2 \pi  u l )-2 \pi  u \sin (2 \pi  u l)\right]\right|.
\end{align}
In the second equality we have used $l = \sin\phi$ for $m=0$.

\end{appendices}

\end{document}